\documentclass[final,3p,times,twocolumn,authoryear]{elsarticle}

\usepackage{amssymb}

\usepackage{graphicx,tabularx}
\usepackage{upgreek}
\usepackage{multicol,multirow}
\usepackage{amsmath,amssymb,amsfonts}
\usepackage{mathrsfs}
\usepackage{amsthm}
\usepackage[figuresright]{rotating}
\usepackage{appendix}
\usepackage{ifpdf}
\usepackage[T1]{fontenc}
\usepackage{newtxtext}
\usepackage{newtxmath}
\usepackage{textcomp}
\usepackage{xcolor}
\usepackage[colorlinks,allcolors=blue]{hyperref}
\definecolor{jourcolor}{cmyk}{1,0.57,0.01,0.38}
\hypersetup{
    colorlinks,%
    citecolor=jourcolor,%
    filecolor=jourcolor,%
    linkcolor=jourcolor,%
    urlcolor=jourcolor
}

\theoremstyle{definition}

\definecolor{ForestGreen}{RGB}{34,139,34}

\journal{International Journal of Multiphase Flow}

\begin{document}

\begin{frontmatter}



\title{A simulation modeling framework for fluid motion and transport in a rocking bioreactor with application to cultivated meat production}


\author[label1]{Minki Kim}
\ead{minki_kim@brown.edu}
\author[label1]{Daniel M. Harris}
\ead{daniel_harris3@brown.edu}
\author[label2]{Radu Cimpeanu\corref{cor1}}
\ead{Radu.Cimpeanu@warwick.ac.uk}

\cortext[cor1]{Corresponding author}
\affiliation[label1]{organization={School of Engineering, Brown University},
            addressline={}, 
            city={Providence},
            postcode={02912}, 
            state={Rhode Island},
            country={USA}}
\affiliation[label2]{organization={Mathematics Institute, University of Warwick},
            addressline={}, 
            city={Coventry CV4 7AL},
            postcode={02912}, 
            state={},
            country={United Kingdom}}

\begin{abstract}
Rocking or wave-mixed bioreactors have emerged as a promising innovation in the production of cultivated meat due to their disposable nature, low operating costs, and scalability. However, despite these advantages, the performance of rocking bioreactors is not well characterized in view of their relatively short history in the market and the wide range of geometrical and operating parameters. In the present study we develop a rigorous computational framework for this multiphase multi-physics system to quantitatively evaluate mixing, oxygen transfer, and shear stress within a rectangular rocking bioreactor under various operating conditions. This framework is implemented using the Basilisk open-source platform. We use a second-order finite volume Navier-Stokes solver and a volume-of-fluid interface reconstruction scheme to accurately resolve the highly nonlinear fluid motion. 
By solving the advection-diffusion equation for a multi-fluid system, we examine mixing time and the oxygen mass transfer coefficient ($k_La$) for different operating conditions, both of which show a strong relationship with steady streaming underlying the instantaneous laminar flow.
We further highlight two critical hydrodynamic phenomena that significantly influence bioreactor performance. Firstly, we investigate the transitional regime from laminar to turbulent flow. Moreover, we identify specific operating conditions that trigger resonance within the bioreactor, enhancing mixing and oxygen transfer. We finally discuss the potential effects of shear stress and energy dissipation rate on cell survival. Our findings are expected to provide valuable insights and guidelines for designing optimized bioreactors to support the next-generation cultivated meat industry pipelines.
\end{abstract}








\end{frontmatter}



\section{Introduction} \label{sec:intro}

Conventional livestock production has been recognized as a primary contributor to habitat and biodiversity loss \citep{Machovina2015,Zabel2019}, as well as water and nutrient pollution \citep{Poore2018} and food-related greenhouse gas emissions \citep{Poore2018,Xu2021}. Their environmental impacts are recognized as significant drivers of climate change and threats to the sustainability of our planet. In response, cultivated meat, also known as lab-grown or cell-based meat, has emerged as a promising and environmentally friendly alternative to traditional meat production \citep{Santo2020}. In 2023, the United Nations highlighted the growing demand for animal source foods, underscoring their potential to mitigate environmental impacts, reduce carbon dioxide emissions, and address animal welfare concerns \citep{UN2023}. A significant milestone was achieved in November 2022 when the first regulatory approval for cultivated meat was granted in the United States. Following this, the United States Food and Drug Administration approved Upside Foods for its cultivated chicken, which became commercially available in June 2023 \citep{FDA2023}. This regulatory approval is expected to accelerate the integration of sustainable alternatives, such as cultivated meat, into the meat production industry. While these advancements mark significant progress, technical and economic challenges remain before such alternatives can become a mainstream commodity \citep{Humbird2021}. One of the primary barriers is scaling up production to an industrial level, which is constrained by several factors, including slow cell growth rates, low final cell densities due to catabolic inhibition, and the high capital costs associated with aseptic operation. Addressing these challenges through continued innovation and process optimization will be critical to fully realizing the potential of cultivated meat within the food industry.

In the manufacturing process of cultivated meat, bioreactors play a crucial role in cultivating animal cells in an oxygen-rich culture medium and facilitating cell differentiation to form muscle tissue. The cultivated cells are then harvested and processed into final meat products. Various types of bioreactors have been employed \citep{Rivera2022}, including stirred tank \citep{Liu2012,Bach2017,Wodolazski2020,Kahouadji2022}, rocking or wave-mixed \citep{Singh1999,Eibl2010,Junne2013}, airlift \citep{Mashhadani2015,Mutaf2023}, hollow fiber \citep{Mohebbi2012}, and rotary wall types \citep{Cinbiz2010}. They are used not only for animal cell cultivation but also for applications such as algae growth \citep{Duan2014}, microbial fermentation \citep{Gill2008}, and stem cell expansion \citep{King2007}. Among these, rocking or wave-mixed bioreactors have emerged as a particularly promising innovation owing to their disposable nature, cost-effectiveness, and scalability. In 1999, \citet{Singh1999} first introduced a rocking or wave-mixed bioreactor system, demonstrating cultivation in volumes up to 100 L. The use of a disposable cellbag as the primary culturing structure allows for quick setup and flexibility in establishing new production facilities \citep{Eibl2010,Junne2013}. In addition, by inducing gentle liquid motions, rocking bioreactors reduce shear damage \citep{Singh1999,Junne2013}, in contrast to other bioreactor types such as stir-mixed and airlift bioreactors. These conventional systems rely on mechanical mixers and sparger-induced bubbles, which pose a higher risk of causing damage to cultured cells. 

In the design of rocking bioreactors, several key factors--oxygen transfer, mixing, shear stress, and energy dissipation rate--play a critical role in cell cultivation. Aerobic microorganisms, such as mammalian cells, rely on oxygen for their metabolic activities, making it essential to maintain sufficient oxygen levels in the medium to support cell growth. Efficient mixing is crucial for the uniform distribution of oxygen, nutrients, temperature, and even cells within the medium, allowing for continuous cell cultivation. However, excessive shear stress and high energy dissipation rates can cause physical damage or alterations in cell behavior, particularly in delicate cells, such as mammalian cells. Therefore, a careful balance of oxygen supply, mixing, shear stress, and energy dissipation rate are necessary to create a healthy and effective cultivation environment while reducing the risk of cell damage.

Several experimental studies have investigated methodologies for accurately characterizing rocking bioreactor performance. Mixing time and oxygen mass transfer rates were experimentally measured using a fluorescent tracer dye and an oxygen-sensitive dye to visually assess mixing and dissolved oxygen in water, a technique referred to as the ``classic dynamic method'' \citep{Singh1999}. More recently, \citet{Bai2019a} employed the acidic tracer method, introducing a pH solution into the medium and evaluating solution homogeneity by monitoring local pH changes with a probe. They also used a dissolved oxygen probe to measure local oxygen concentrations, similar to \citet{Singh1999}. \citet{Bartczak2022} utilized local pH measurements to determine mixing time. While these studies have quantified mixing and oxygen supply metrics using tracers, the dependence on point-wise measurements introduces significant uncertainty due to the limited information regarding the overall tracer distribution within the medium. Furthermore, nearly all empirical scaling relations derived for mixing time and oxygen transfer rate under given operating conditions are tailored to turbulent flows. Thus, directly extending these findings to laminar and transitional flow regimes is not straightforward, as the key advective transport mechanisms can be rather different. As an alternative to point-wise measurements, \citet{Marsh2017} conducted phase-resolved Particle Image Velocimetry (PIV) experiments to measure flow characteristics across the cross-section of the cellbag for the first time. However, their study focused only on flow velocity visualization, which limits the ability to comprehensively evaluate bioreactor performance.

As a complementary effort, computational approaches have provided valuable insights into the flow characteristics of rocking bioreactors. Specifically, shear stress in the medium can be estimated using flow velocity and vorticity, which are readily accessible in simulations, offering key information about potential adverse effects to cultured cells. Pointwise measurements of velocities and shear stress have been quantified using commercial software, aiding in the understanding of mixing and potential cell damage, respectively \citep{Oncul2010,Zhan2019}. Similarly, pointwise oxygen concentration measurements have been used to quantify oxygen mass transfer rates. In contrast, \citet{Svay2020} and \citet{Seidel2022} quantified overall flow velocity within bioreactors and used it to estimate oxygen mass transfer rates based on existing scaling relations. While simulation results provide a more accurate representation of shear stress and offer nearly continuous velocity data over time, there remains a need for advanced numerical capabilities to comprehensively assess bioreactor performance. However, this is particularly challenging due to the complex geometrical configurations and multi-physics aspects of the flow.

In this study, we develop a computational framework for simulating rocking bioreactors using the Basilisk open-source platform (\href{http://basilisk.fr}{http://basilisk.fr}) \citep{Popinet2003,Popinet2009,Popinet2015}. By employing a non-inertial frame of reference, we accurately capture the rocking motions while maintaining stationary boundaries for the cellbags. Furthermore, we adapted the capabilities of the platform to account for oxygen transfer across different phases, enabling a more robust evaluation of mixing and oxygen supply directly from first principles. Using this framework, we investigate the flow characteristics of a rectangular-shaped bioreactor under various operating conditions, aiming to elucidate their relation with overall bioreactor performance, including metrics such as mixing time, oxygen mass transfer coefficient, shear stress, and energy dissipation rate. The implementation is publicly available for interested users at \href{https://github.com/rcsc-group/BioReactor}{https://github.com/rcsc-group/BioReactor}.

\section{Methods} \label{sec:method}

\subsection{Numerical framework}
In the following sections, we elaborate on the methodological building blocks of our approach, including an introduction to the computational platform used in this study, its adaptations, and the key mathematical modeling techniques that have significantly enhanced its applicability and efficiency.

\subsubsection{Basilisk solver}

We solve the two-dimensional, incompressible, two-phase Navier-Stokes equations along with the advection equation for volume fraction using an implementation in the open-source platform Basilisk \citep{Popinet2003,Popinet2009,Popinet2015}. The equations are as follows:
\begin{equation}
    \begin{split}
        \rho \frac{\mathrm{D} \mathbf{u}}{\mathrm{D} t} = -\nabla p + & \nabla \cdot \left( 2\mu\mathbf{D}\right) + \sigma\kappa\delta_s\mathbf{n} + \mathbf{f}, \\
        \nabla \cdot \mathbf{u} & = 0, \\
        \frac{\partial \alpha}{\partial t} + \mathbf{u} & \cdot \nabla \alpha = 0, \label{eq:NS-inertial}
    \end{split}
\end{equation}
where $\rho$ is the density, $\mathrm{D}$ is the material derivative, $\mathbf{u}$ is the velocity, $p$ is the pressure, $\mu$ is the viscosity, $\mathbf{D}$ is the deformation tensor, $\sigma$ is the surface tension, $\kappa$ is the curvature, $\delta_s$ is the Kronecker delta function restricting the term in question at the interface, $\mathbf{n}$ is the unit vector normal to the interface, $\mathbf{f}$ is the body force per unit volume, and $\alpha$ is the volume fraction of water. For two-phase flows, the density and viscosity are defined as $\rho = \alpha\rho_w + (1-\alpha)\rho_a$ and $\mu = \alpha\mu_w + (1-\alpha)\mu_a$, where $w$ and $a$ refer to the water and the air, respectively. It is worth noting that water properties are used for the liquid phase in the present work; however, the formulation is relevant to any Newtonian fluid, and can be extended to account for more complex fluid properties in future applications, such as when simulating cell culturing. The second-order finite volume Navier-Stokes solver is used, and a volume-of-fluid interface reconstruction scheme is implemented to resolve the highly nonlinear fluid motion accurately. The Basilisk infrastructure, extensively validated by previous studies \citep{Ojiako2020,Farsoiya2021,Li2021,Hidman2022,Fudge2023,Alventosa2023,Radu2023,Xu2025}, has been widely employed for multiphase and interfacial flow problems, including droplet impact \citep{Fudge2023,Alventosa2023}, liquid-gas turbulent flow \citep{Farsoiya2021,Farsoiya2023}, needle design \citep{Radu2023}, bubble dynamics \citep{Li2021,Hidman2022}, and liquid film \citep{Ojiako2020} studies. With its capacity to appropriately handle gas-liquid flow within confined geometries, Basilisk is well-suited for modeling a rocking bioreactor. In addition to its inherent capabilities, we integrate features to account for the non-inertial frame of reference, ensuring stationary boundaries. This integration significantly reduces computational costs compared to bioreactor modeling in an inertial (lab) frame of reference, where the bioreactor geometry oscillates within the computational domain. Furthermore, we leverage Basilisk's capability for mass transfer modeling to consider oxygen transfer and to assess mixing. These features will be detailed in the following subsections.

\subsubsection{Governing equations in a non-inertial frame of reference}

\begin{figure}[t!]
\center \includegraphics[width=0.48\textwidth]{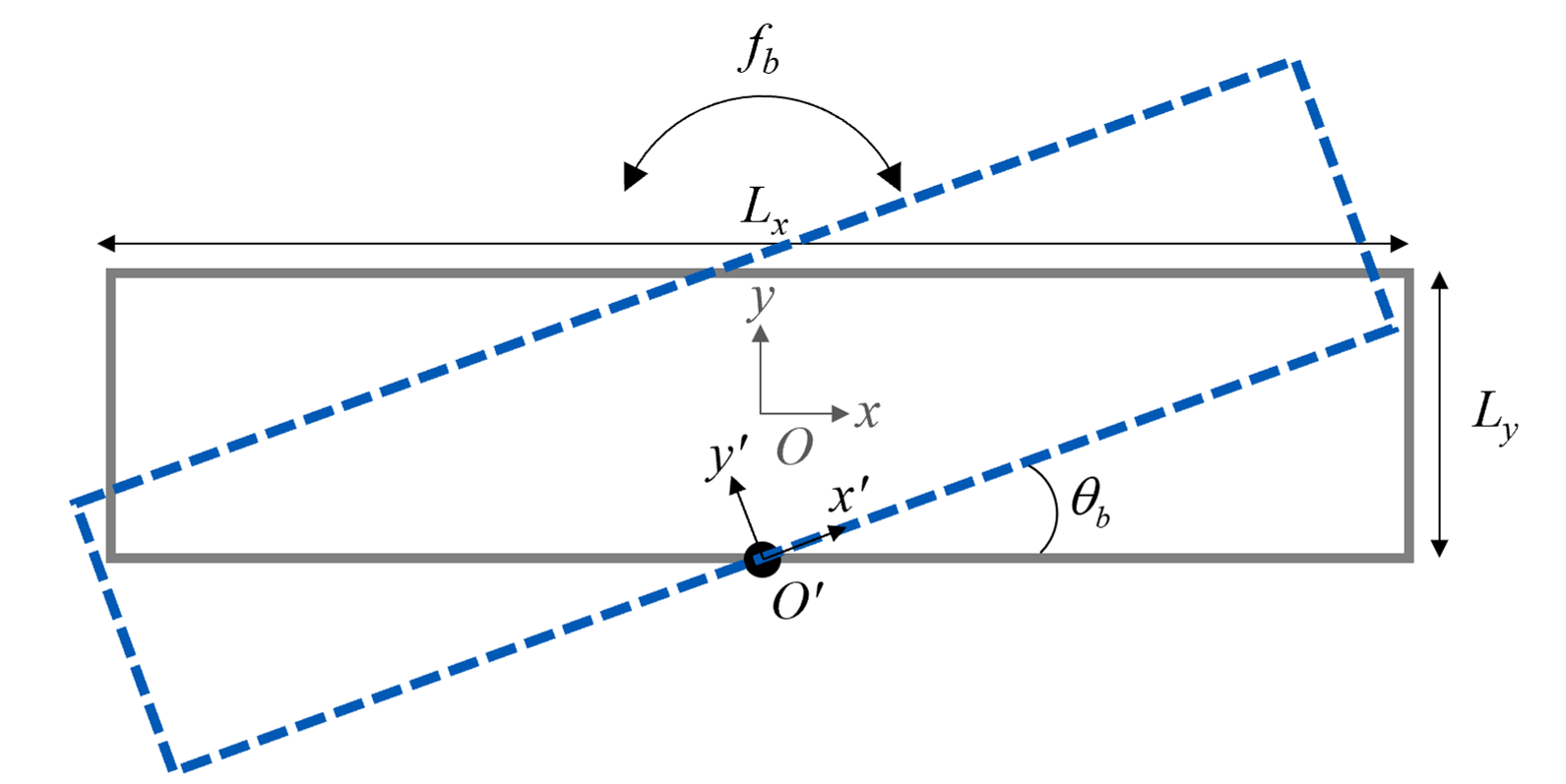}
\caption{Inertial $O$ (gray solid rectangle) and non-inertial $O'$ (blue dashed rectangle) frame of reference for a rocking (or wave-mixed) bioreactor. The width $L_x$ and height $L_y$ of the bioreactor geometry with a rocking angle $\theta_b$ and frequency $f_b$. The pivot point, located at the middle of the bottom edge, serves as the axis of rotation for the rocking motion.} \label{Fig:bioreactor}
\end{figure}

To model the rocking motion of the bioreactor, it is crucial to consider gas-liquid flows within the no-slip boundaries integrated into the computational domain. These embedded boundaries dynamically follow the rocking motion, presenting numerical and computational challenges. This includes addressing large computational domains covering the bioreactor geometry, resulting in substantial computational loads. Moreover, managing time-dependent wetting boundary conditions becomes important, requiring the consideration of normal and tangent vectors of the boundaries \citep{Sanjeevi2018}. Prior studies have employed the immersed boundary method to account for fluid-solid interaction \citep{Peskin1972}, primarily focusing on single-phase flows \citep{Niu2022}. Recent studies have extended this approach to multiphase flow systems, incorporating curved boundaries \citep{Niu2022} and complex geometries \citep{Patel2017}, which are stationary, as well as moving rigid boundaries \citep{Patel2018,Xiao2022}, which are still under development for accessible and user-friendly platforms.

In the present simulations, a non-inertial frame of reference fixed on the simplified bioreactor geometry defines the computational domain. This frame, along with an inertial frame of reference and bioreactor geometry, is illustrated in Fig. \ref{Fig:bioreactor}. The inertial frame of reference $O$ remains stationary as the geometry exhibits no translational motion. On the other hand, the non-inertial frame of reference $O'$ rotates with a rotational velocity of $\mathbf{\Omega}$ along with the bioreactor geometry. The pivot point, located at the middle of the bottom edge, serves as the axis of rotation for the rocking motion.

For a fluid particle $P$, the velocity $\mathbf{u}'$ viewed in the rotating frame $O'$ is expressed as $\mathbf{u}' = \mathbf{u} - \mathbf{U} - \mathbf{\Omega} \times \mathbf{r}'$ \citep{Kundu2015}, where $\mathbf{u}$ is the particle velocity viewed in the inertial frame $O$, $\mathbf{U}$ is the translational velocity of the rotating frame $O'$ relative to $O$, and $\mathbf{r}'$ is the particle position in the frame $O'$. We note that $\mathbf{U}=0$ in the present study. Correspondingly, the particle acceleration $\mathbf{a}'$ viewed in the rotating frame $O'$ is given by $\mathbf{a}' = \mathbf{a} - [\dot{\mathbf{U}} + 2\mathbf{\Omega}\times\mathbf{u}' + \dot{\mathbf{\Omega}}\times\mathbf{x}' + \mathbf{\Omega}\times(\mathbf{\Omega}\times\mathbf{x}')]$, where $\mathbf{a}$ is the particle acceleration in the inertial frame $O$, the first term in the square brackets is the acceleration of frame $O'$ relative to $O$, the second term is the Coriolis acceleration, the third term is the angular acceleration, and the last term is the centrifugal acceleration. Utilizing velocity and acceleration relations for the frame transformation, governing equations \eqref{eq:NS-inertial} in the rotating frame $O'$ become:
\begin{equation}
    \begin{split}
        \rho\frac{\mathrm{D}' \mathbf{u}'}{\mathrm{D} t} = -\nabla' p + & \nabla' \cdot \left( 2\mu\mathbf{D}'\right) + \sigma\kappa'\delta_s'\mathbf{n}' + \mathbf{f'}, \\
        \nabla' \cdot \mathbf{u}' & = 0, \\
        \frac{\partial' \alpha}{\partial t} + \mathbf{u}' & \cdot \nabla' \alpha = 0, \label{eq:NS-rot}
    \end{split}
\end{equation}
where $'$ denotes the quantity viewed in the frame $O'$, $\mathbf{f'} = \rho [\mathbf{g} - (2\mathbf{\Omega}\times\mathbf{u}' + \dot{\mathbf{\Omega}}\times\mathbf{x}' + \mathbf{\Omega} \times (\mathbf{\Omega} \times \mathbf{x}'))]$, and $\mathbf{g}$ is the gravitational acceleration in the vertical direction viewed in the inertial frame $O$. We note that only the momentum conservation equation has additional terms referred to as fictitious forces \citep{Kundu2015,Gledhill2016,Combrinck2017,Musehane2021}. The continuity and scalar advection equations remained unchanged, consistent with previous studies, where mass conservation equation \citep{Gledhill2016,Combrinck2017} and scalar values \citep{Gledhill2016,Combrinck2017,Musehane2021} remain unaffected by the transformation of the frame. Thus, the Basilisk solver, originally designed for solving Eq. \eqref{eq:NS-inertial}, can be directly applied to solve Eq. \eqref{eq:NS-rot}, taking into account the fictitious forces by treating them as body forces in the modified equations.

\begin{figure*}[t!]
\center \includegraphics[width=0.9\textwidth]{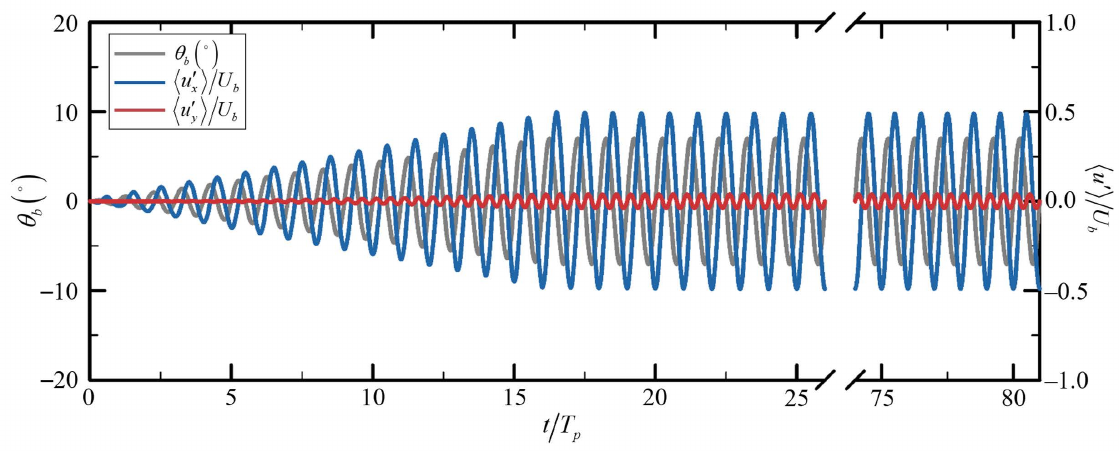}
\caption{Rocking profile for the maximum rocking angle $\theta_{b,\textrm{max}}=7^{\circ}$ and $f_b=32.5$ rpm. The time evolution of the rocking angle (gray solid lines) and the spatially averaged $x'$ (blue solid lines) and $y'$ (red solid lines) velocities, normalized by the characteristic velocity scale in the non-inertial frame of reference, are shown during the early stage of oscillation and near the time instant ($t/T_p$ = 80) when the tracer and oxygen are introduced.} \label{Fig:simul_setup}
\end{figure*}

\subsubsection{Volume-of-Fluid method for interfacial mass transfer}

The computational approach for modeling mass transfer across interfaces in multiphase flow systems in the framework of the Volume-of-Fluid (VoF) method was initially developed by \citet{Haroun2010}. This method was integrated into Basilisk for conducting direct numerical simulations of a dissolving gas bubble in turbulent flows \citep{Farsoiya2021,Farsoiya2023}. In the present study, this recent functionality has been employed to account for mass transfer, facilitating the assessment of mixing and oxygen transfer in the bioreactor. The overall methodologies are introduced as follows.

The conservation equation governing the local concentration $C_{k,j}$ of the chemical species $j$ in phase $k$ is expressed as follows:
\begin{equation}
    \frac{\partial C_{k,j}}{\partial t} + \nabla' \cdot \left( \mathbf{u}'_k C_{k,j}\right) = -\nabla' \cdot \mathbf{J}'_{k,j}, \label{eq:C_local}
\end{equation}
where $k$ denotes water ($w$) or air ($a$), and $j$ denotes oxygen (oxy) and tracer (tr). The molecular flux $\mathbf{J}'_{k,j} = -D_{k,j}\nabla C_{k,j}$ follows Fick's law, and $D_{k,j}$ is the diffusivity of the chemical species $j$ in phase $k$. 

At the interface $\Sigma$, the normal component of the flux $\mathbf{J}'_{k,j}$ is continuous, satisfying $\mathbf{J}'_{w,j}\cdot \mathbf{n}'_\Sigma = \mathbf{J}'_{a,j} \cdot \mathbf{n}'_\Sigma$. The dissolution equilibrium of the chemical species is achieved through Henry's law: $C_{w,j}=C_{a,j}H_j$, implying discontinuity in the concentration fields across the interface if the solubility constant $H_j \neq 1$. Utilizing the one-fluid formulation with the VoF method, where $C_j = \alpha C_{w,j} + (1-\alpha)C_{a,j}$ and $\mathbf{J}'_j = \alpha\mathbf{J}'_{w,j} + (1-\alpha)\mathbf{J}'_{a,j}$, \citet{Haroun2010} and \citet{Farsoiya2021} derived a unified conservation equation for both phases:
\begin{equation}
    \frac{\partial C_j}{\partial t} + \nabla' \cdot \left( \mathbf{u}' C_j \right) = \nabla' \cdot \left(D_j\nabla' C_j + \mathbf{\Phi}'_j \right), \label{eq:C_one-fluid}
\end{equation}
where $\mathbf{\Phi}'_j = - D_j C_j(H_j-1)\nabla' \alpha/(H_j\alpha + 1-\alpha)$ is introduced due to the discontinuous concentration following Henry's law. Importantly, under the one-fluid formulation, the diffusivity at the interface is determined using the harmonic mean of the diffusivity of each phase: $1/D_j = \alpha/D_{w,j} + (1-\alpha)/D_{a,j}$. In the present study, Eq. \eqref{eq:C_one-fluid} was employed to model the transfer of oxygen from air to water and the advection and diffusion of oxygen in each phase. In addition, advection and diffusion of the tracer, released within the water, were considered for mixing evaluation using Eq. \eqref{eq:C_one-fluid}. In the context of infinite solubility ($H_j \rightarrow \infty$), this confines the tracer within the water without any transfer into the air.

\subsection{Problem description and simulation setup}

\begin{table*}[h]
\centering
\caption{Relevant quantities, notation, and values for all the relevant parameters used in this investigation.} \label{tab:sim_param}
\begin{tabular}{m{15em} c c c}
\hline
Parameter & Symbol & Definition & Value \\ \hline
Domain width & 
$L_x$ & 
--- & 
0.250 m \\
Domain size aspect ratio & 
$\beta_b$ & 
$L_y/L_x$ & 
0.285 \\
Maximum rocking angle & 
$\theta_{b,\textrm{max}}$ & 
--- & 
2--7$^{\circ}$ \\
Rocking frequency & 
$f_b$ & 
--- & 
15--37.5 rpm \\
Rocking period & 
$T_p$ & 
--- & 
1.6--4.0 s \\
Characteristic velocity scale & 
$U_b$ & 
$L_x(1+\mathrm{tan}\theta_{b,\textrm{max}}/2\beta_b)/2T_p$ & 
0.038--0.095 m/s \\
\hline
Density (water)  & 
$\rho_w$ & 
--- & 
1.00 $\times$ 10$^3$  kg/m$^3$ \\
Viscosity (water)  & 
$\mu_w$ & 
--- & 
1.00 $\times$ 10$^{-3}$ Pa $\cdot$ s \\
Diffusivity of tracer (water)  & 
$D_{w,\textrm{tr}}$ & 
--- & 
4.40 $\times$ 10$^{-10}$ m$^2$/s \\
Diffusivity of oxygen (water)  & 
$D_{w,\textrm{oxy}}$ & 
--- & 
1.90 $\times$ 10$^{-9}$ m$^2$/s \\
Density (air)  & 
$\rho_a$ & 
--- & 
1.225 kg/m$^3$ \\
Viscosity (air)  & 
$\mu_a$ & 
--- & 
1.81 $\times$ 10$^{-5}$ Pa $\cdot$ s \\
Diffusivity of tracer (air)  & 
$D_{a,\textrm{tr}}$ & 
--- & 
1.00 $\times$ 10$^{-16}$ m$^2$/s \\
Diffusivity of oxygen (air)  & 
$D_{a,\textrm{oxy}}$ &
--- & 
1.98 $\times$ 10$^{-5}$ m$^2$/s \\
Gravitational acceleration  & 
$g$ & 
--- & 
9.8 m/s$^2$ \\
Surface tension coefficient  & 
$\sigma$ & 
--- & 
0.0728 N/m \\
Solubility constant (tracer)  & 
$H_{\textrm{tr}}$ & 
--- & 
1.00 $\times$ 10$^{16}$ \\
Solubility constant (oxygen)  & 
$H_{\textrm{oxy}}$ & 
--- & 
0.0333 \\
\hline
Reynolds number (water) & 
$Re_w$ & 
$\rho_w U_b L_x/\mu_w$ & 
9.5$\times 10^3$--2.4$\times 10^4$ \\
Reynolds number (air) & 
$Re_a$ & 
$\rho_a U_b L_x/\mu_a$ &
650--1600 \\
Weber number (water) & 
$We_w$ & 
$\rho_wU_b^2L_x/\sigma$ & 
5--31 \\
Weber number (air) & 
$We_a$ & 
$\rho_aU_b^2L_x/\sigma$ & 
0.006--0.038 \\
Bond number (water) & 
$Bo_w$ & 
$\rho_wgL_x^2/\sigma$ & 
8,400 \\
Bond number (air) & 
$Bo_a$ & 
$\rho_agL_x^2/\sigma$ & 
10.3 \\
Peclet number of tracer (water) & 
$Pe_{w,\textrm{tr}}$ & 
$U_bL_x/D_{w,\textrm{tr}}$ & 
2.2$\times 10^7$--5.4$\times 10^7$ \\
Peclet number of tracer (air) & 
$Pe_{a,\textrm{tr}}$ & 
$U_bL_x/D_{a,\textrm{tr}}$ & 
9.5$\times 10^{13}$--2.4$\times 10^{14}$  \\
Peclet number of oxygen (water) & 
$Pe_{w,\textrm{oxy}}$ & 
$U_bL_x/D_{w,\textrm{oxy}}$ & 
5.0$\times 10^6$--1.25$\times 10^7$ \\
Peclet number of oxygen (air) & 
$Pe_{a,\textrm{oxy}}$ & 
$U_bL_x/D_{a,\textrm{oxy}}$ & 
480--1,200 \\
Froude number & 
$Fr$ &
$U_b/\sqrt{gL_x}$& 
0.024--0.061 \\
\hline
\end{tabular}
\end{table*}

A series of bioreactor simulations with a rectangular fixed domain were conducted under different operating conditions: a maximum rocking angle $\theta_{b,\textrm{max}}$ and a rocking frequency $f_b$ in the ranges of $2^{\circ} \leq\theta_{b,\textrm{max}}\leq 7^{\circ}$ and  $15 \leq f_b \leq 37.5$ rpm. Table \ref{tab:sim_param} summarizes the ranges of simulation parameters and associated dimensionless numbers. The selected rectangular geometry has a width $L_x$ of 0.25 m and an aspect ratio $\beta_b$ of 0.285, which generally reflects the shape of high-aspect-ratio 2L cellbags. This single geometry can broadly represent a variety of cellbag shapes and dimensions, including those produced by Cell-tainer \citep{Junne2013}, Cytiva \citep{Singh1999,Bartczak2022}, BioWave \citep{Eibl2010}, and Wave Biotech AG \citep{Oncul2010,Kalmbach2011}. Typically, these cellbags have a larger width than height $L_y$, which enhances the exposure of oxygen in the air to the water. The ranges of operating conditions were selected based on the rocking angles and frequencies that result in gentle mixing within the water, indicative of the laminar flow regime. These ranges also align with operating conditions used in previous studies \citep{Oncul2010,Kalmbach2011,Marsh2017}.

In the present study, the rocking motion of the bioreactor is simulated using a sinusoidal periodic motion \citep{Zhan2019}, expressed as $\theta_b(t) = \theta_{b,\textrm{max}} \mathrm{sin}(\omega_b t)$, where $\omega_b = 2\pi f_b$ is the angular frequency. A sample rocking motion for the representative scenario is shown in Fig. \ref{Fig:simul_setup}. The amplitude of the rocking motion linearly increases until it reaches a regular amplitude, avoiding abrupt acceleration at the start, especially near the domain boundary, which initially has zero velocity, and mimicking practical bioreactor motions. This constant amplitude at the regular mode is attained after 30 seconds, ensuring a gradual acceleration pattern; all subsequent analyses are conducted after this transient period.

For the analysis, we selected $\theta_{b,\textrm{max}} = 7^\circ$ and $f_b=32.5$ rpm as a representative scenario. This condition induces gentle mixing with nonlinear liquid motion, exhibiting distinct flow characteristics. In these operating conditions, the horizontal and vertical velocities, spatially averaged over the water phase ($\langle u_x'\rangle$ and $\langle u_y'\rangle$, respectively) and normalized by the characteristic velocity scale $U_b$, are illustrated together with the rocking angle in Fig. \ref{Fig:simul_setup}. These velocities exhibit periodic behavior and closely correspond to the rocking angle, showcasing a laminar flow regime. The reference velocity scale $U_b=L_x(1+\mathrm{tan}\theta_{b,\textrm{max}}/2\beta_b)/2T_p$ is determined as the horizontal velocity calculated using the volumetric rate from one half of the geometry to the other, divided by the vertical length in the mid-plane \citep{Marsh2017}. In simulations, half of the geometry is filled with water, while the remaining half is occupied by air, representing a 1L liquid volume for the 2L cellbag. The reference length scale is taken to be $L_b=L_x$, which governs both the mixing within the water and oxygen transfer across the interface. A uniform mesh with $n_L=2^{10}$ grid cells along the width of the domain is used. Each grid cell has a dimensional size of 0.24 mm, resulting in a total of $1.05 \times 10^6$ cells. This configuration requires 120 cores and 120 CPU core hours to account for both mixing and oxygen transfer under $\theta_{b,\textrm{max}} = 7^\circ$ and $f_b=32.5$ rpm. Further details on the convergence of spatially averaged $x'$ and $y'$ velocities in a bioreactor are provided in \ref{app:grid}. 

The properties of water and air at room temperature were employed in the simulations for the sake of the current exploration. The diffusivity of the tracer $D_{w,\textrm{tr}}$ in water corresponds to that of rhodamine, a commonly used tracer in previous studies \citep{Bach2017,Rodriguez2018,Atanasova2023}. Meanwhile, the diffusivity of the tracer in air $D_{a,\textrm{tr}}$ is assumed to be effectively zero, thus neglecting possible tracer diffusion within the air. In addition, since the tracer is present only in the water in the real experiment, its solubility constant is set to be a very large value in the simulations, effectively preventing tracer transport across the interface. We define dimensionless numbers in the bioreactor system based on the properties of water and air, and the characteristic length and velocity scales. The ranges of the Reynolds number $Re=\rho U_b L_x/\mu$, Weber number $We=\rho U_b^2 L_x /\sigma$, Bond number $Bo=\rho gL_x^2/\sigma$, Peclet number $Pe=U_bL_x/D$, and Froude number $Fr = U_b/\sqrt{gL_x}$ are specified in Table \ref{tab:sim_param}. We especially note the Peclet number of oxygen being approximately $O(10^{6})$ in water and $O(10^{2})$ in air, indicating that fluid motion primarily induces oxygen species transport as opposed to molecular diffusion. The Bond number for water is 8,400 for the chosen geometry, indicating that the interfacial waves are gravity-driven.

\section{Evaluation of bioreactor performance} \label{sec:performance}

\subsection{Steady streaming in a bioreactor} \label{subsec:steady_stream}

\begin{figure}[t!]
\center \includegraphics[width=0.48\textwidth]{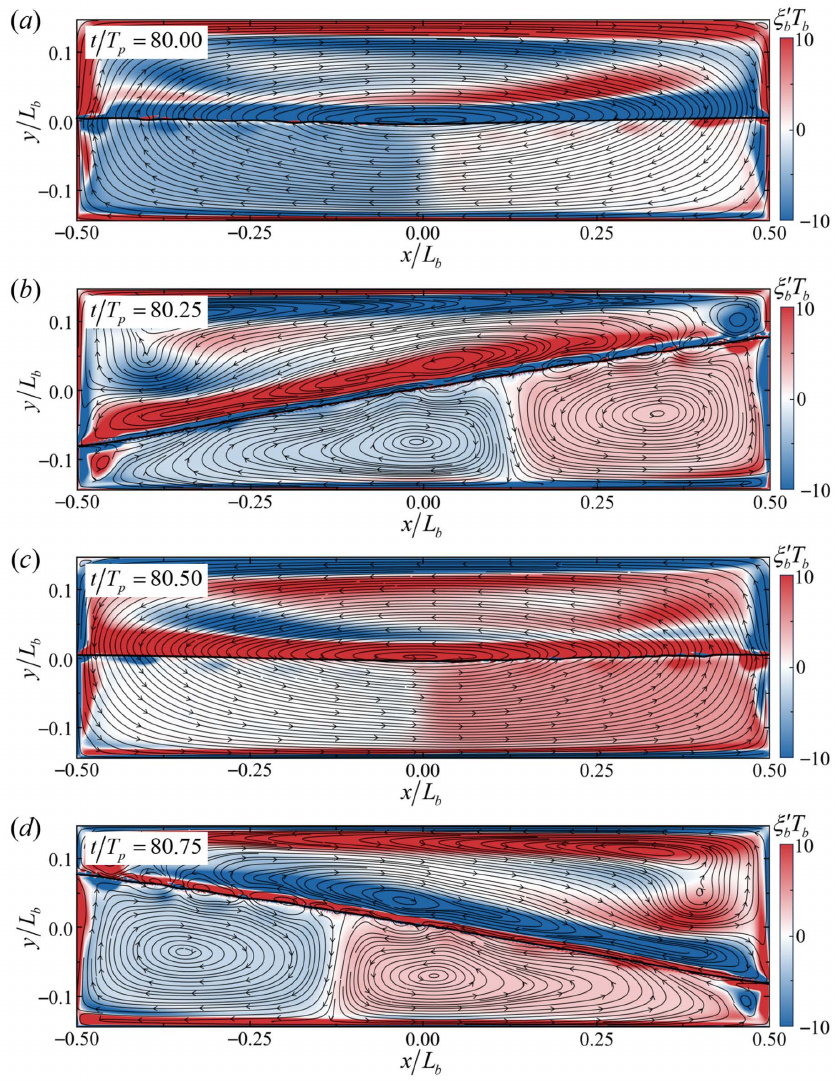}
\caption{Instantaneous vorticity fields overlaid with streamlines in a bioreactor for $\theta_{b,\textrm{max}}= 7^{\circ}$ and $f_b=32.5$ rpm at (a) $t/T_p=$ 80.00, (b) 80.25, (c) 80.50, and (d) 80.75.} \label{Fig:vor_all}
\end{figure} 

The water-air phases and their interface exhibit periodic motion under the current operating conditions, which lie within the laminar flow regime. For the baseline case ($\theta_{b,\textrm{max}}= 7^{\circ}$ and $f_b=32.5$ rpm), two counter-rotating vortical structures emerge at the maximum rocking angles, while horizontal flows dominate during most of the oscillation cycle, as shown in Fig. \ref{Fig:vor_all}. The time-averaged motion of this laminar oscillatory flow induces a net, steady flow--known as steady streaming--which results in the net transport of fluid parcels, as demonstrated in the case of orbital bioreactors in prior work \citep{Bouvard2017}.
During oscillation, the viscous boundary layers along the walls and below the liquid surface create a shear gradient, generating a net force on the flow over a cycle and inducing what is known as steady streaming. 

\begin{figure}[t!]
\center \includegraphics[width=0.48\textwidth]{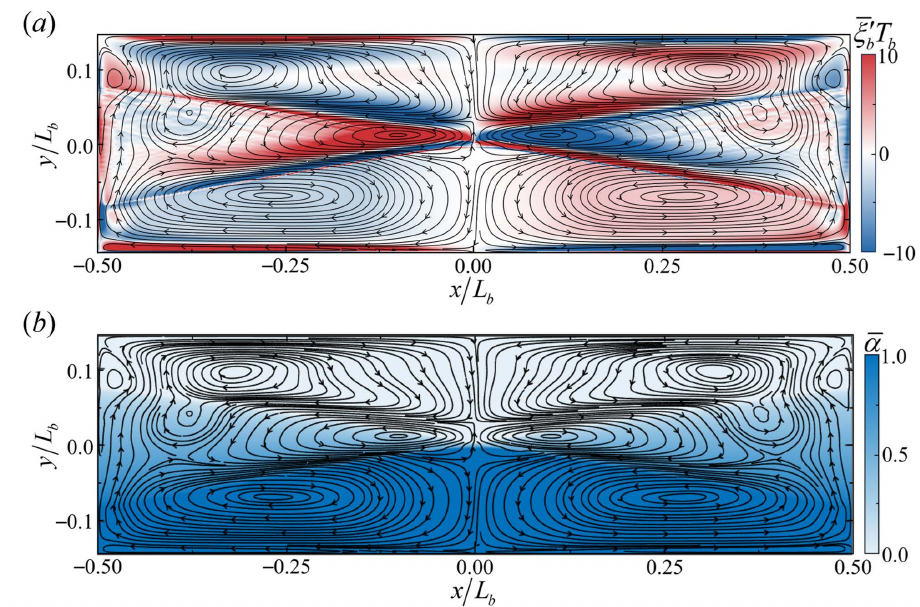}
\caption{Steady streaming in a bioreactor for $\theta_{b,\textrm{max}}= 7^{\circ}$ and $f_b=32.5$ rpm: (a) normalized vorticity field and (b) volume fraction field, each overlaid with streamlines.} \label{Fig:steady_streaming}
\end{figure} 

We obtained steady streaming flow fields by averaging 1,000 instantaneous flow fields over one period. It is worth noting that in the following sections, an averaged flow field is also derived for flows exhibiting turbulent-like (aperiodic) behavior to ensure a consistent comparison with laminar flow regimes, even though steady streaming is less clearly defined in such flows. The snapshots in Fig. \ref{Fig:steady_streaming} depict period-averaged vorticity ($\overline{\xi_b'}$) and volume fraction ($\overline{\alpha}$) fields, overlaid with streamlines of the time-averaged velocity field at $\theta_{b,\textrm{max}} = 7^\circ$, $f_b = 32.5$ rpm. Here, the overbar denotes the period-averaged value for each cell. The flow fields reveal distinct structures based on volume fraction (Fig. \ref{Fig:steady_streaming}(b)): a pure water region ($\alpha=1$) and an air region ($\alpha=0$) at the lower and upper areas of the bioreactor, with an intermediate region ($0 \leq \alpha \leq 1$) characterized by different flow structures (Fig. \ref{Fig:steady_streaming}(a)). Specifically, the liquid side features two large counter-rotating vortical structures, distorted by interfacial motion. In complementary Lagrangian simulations of passive tracer particles (not included in the manuscript), these vortices are confirmed to drive net tracer displacements over each period, following the steady streaming streamlines. This flow configuration is expected to be closely linked to mixing and oxygen transfer which develop over timescales significantly longer than an oscillation period, discussed later in Section \ref{sec:results}. Similar vortical structures have been documented in the context of standing Faraday waves \citep{Perinet2017,Alarcon2020}.


\subsection{Mixing in the water} \label{subsec:mixing}

Mixing in a bioreactor is typically assessed by monitoring the concentration of a tracer at a single point \citep{Bach2017,Bai2019a,Svay2020}, analyzing the spatial distribution of the tracer \citep{Seidel2022}, and tracking distributions of the tracer particles released with certain configurations \citep{Siiria2009,Nguyen2018,Watanabe2022}. Depending on the assessment approaches, the degree of mixing or homogeneity is defined in different ways. In experimental studies, monitoring pH or conductivity with a probe allows for the evaluation of the degree of mixing by comparing temporary states with a perfectly mixed state \citep{Bach2017,Bai2019a}. Similarly, in simulation studies, the degree of mixing is evaluated by comparing temporary tracer concentrations or particle distributions with those in a perfectly mixed state in each cell \citep{Siiria2009,Nguyen2018,Seidel2022,Watanabe2022}. In the present study, we employ tracers initially released with a prescribed configuration to evaluate the variance in their distribution compared to a well-mixed state, thereby determining the degree of mixing. This approach follows the methodology proposed by \citet{Jha2011}:
\begin{equation} \label{eq:mixing_chi}
    \chi = 1-\frac{\sigma^2(t)}{\sigma_{\textrm{max}}^2},
\end{equation}
where $\sigma$ is the variance of the tracer distribution, defined as $\sigma^2 = \langle C_{w,\textrm{tr}}^2 \rangle - \langle C_{w,\textrm{tr}} \rangle^2$. Here, $\langle\rangle$ implies spatial averaging over the water phase. $\sigma_{\textrm{max}}$ denotes a perfectly segregated state when the tracer is initially released ($\chi=0$). Conversely, for a well-mixed state, $\sigma = 0$, yielding $\chi=1$. It is worth noting that tracer advection and diffusion are confined within the water phase by assigning a very large solubility constant on the water phase. This provides a more accurate assessment of mixing compared to methodologies that allow tracers to cross the water-air interface--often due to local approximations of the interface in simulations--which can result in tracer loss from the water phase.

The initial configuration of tracers profoundly affects mixing assessment. \citet{Siiria2009} demonstrated how the primary mixing direction relates to initial tracer configurations, showing that different tracer setups result in varying degrees of mixing regardless of the mixing process. For instance, in a horizontally moving rectangular mixer, tracer particles occupying the top and bottom halves maintain their initial configuration over time, which is more suitable for evaluating mixing in the vertical direction. In addition, the initial tracer setup is critical for mixing assessment in both laminar and turbulent flows. In laminar flows, the performance of the mixer strongly depends on the initial configuration of tracers, such as injection location and flow speeds \citep{Hobbs1997,Zalc2003}. In addition, \citet{Gubanov2009} explored mixing optimization scenarios for different initial tracer fields, crucial for designing mixing devices to achieve uniform mixing quality. In turbulent flows, a recent study by \citet{Singh2023} investigated three different initial conditions of tracers in a channel flow, which are aligned with the streamwise, wall-normal, and traverse directions. The authors found that the fastest mixing occurs when tracers are aligned with the wall-normal direction.

\begin{figure}[t!]
\center \includegraphics[width=0.48\textwidth]{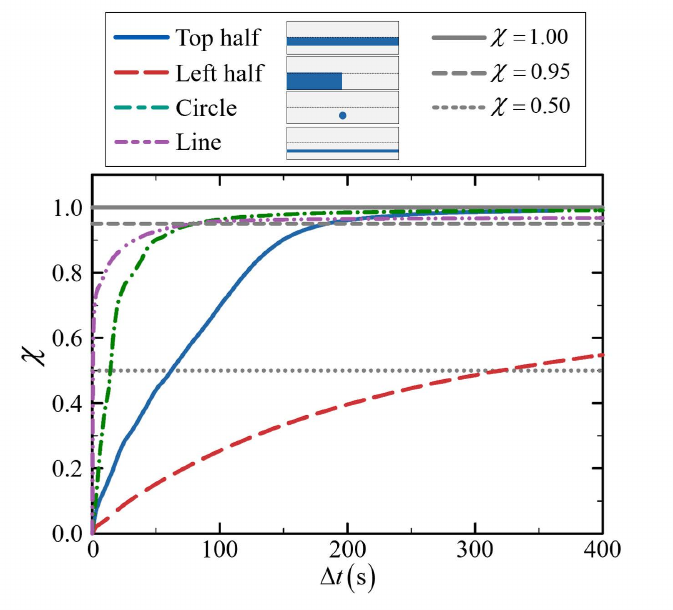}
\caption{The degree of mixing for different initial configurations of tracers for $\theta_{b,\textrm{max}}= 7^{\circ}$ and $f_b=32.5$ rpm: the (blue dashed line) top half and (red dashed-dot line) left half of the water phase and the (green dashed-double-dot line) circular and (orange dashed line) line shapes. Different degrees of mixing are marked: $\chi=1.00$ (perfectly mixed state, gray solid line), $0.95$ (the state defining mixing time, gray dashed line), and $0.50$ (partially mixed state, gray dotted line).} \label{Fig:tracer_ini}
\end{figure}

\begin{figure*}[t!]
\center \includegraphics[width=1.0\textwidth]{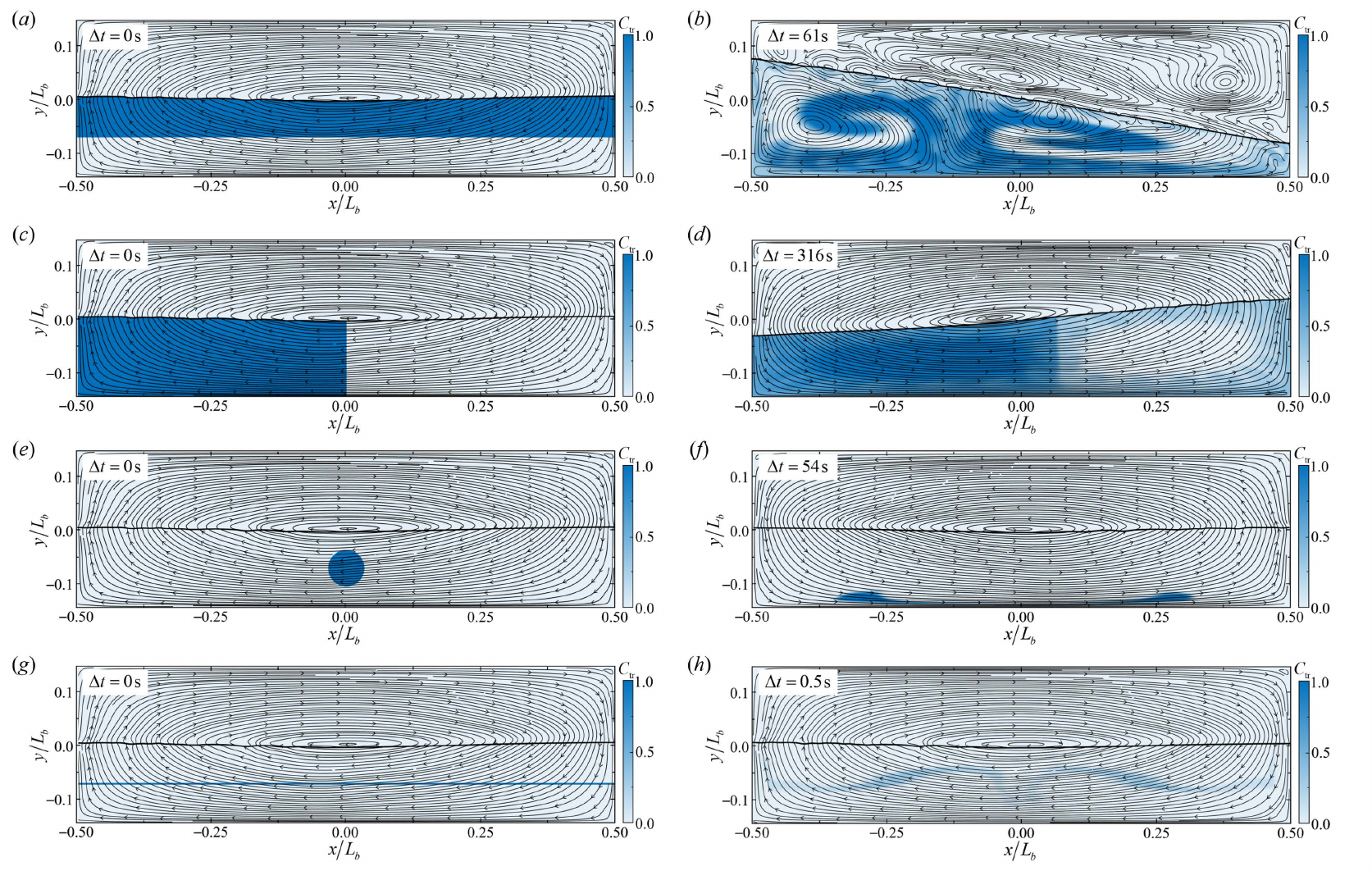}
\caption{Instantaneous tracer fields at (a,c,e,g) $\chi$ =  0 (initial state) and (b,d,f,h) $\chi$ = 0.5 (partially mixed state) for different initial configurations of tracers, marked with a dotted line in Fig. \ref{Fig:tracer_ini}: (a,b) top half and (c,d) bottom half, (e,f) circle at the center and (g,h) line in the middle of the water phase. $\theta_{b,\textrm{max}}= 7^{\circ}$ and $f_b=32.5$ rpm.} \label{Fig:tracer_ini_snap}
\end{figure*}

In the case of a rocking bioreactor, its motion includes both vertical and horizontal components, facilitating efficient mixing and the transfer of oxygen and nutrients within the cellbag. However, initial tracer configurations remain crucial for assessing and quantifying mixing. As a case study, we examined mixing at $\theta_{b,\textrm{max}}=7^{\circ}$ and $f_b=32.5$ rpm for four different tracer configurations: top half, left half, center circle, and middle line in the water phase. The first two scenarios have the same total tracer concentration, as do the last two. The tracers were introduced after 80 cycles, ensuring that the flows exhibited periodic motion with no influence from the initial rocking motion, as illustrated in Fig. \ref{Fig:simul_setup}. The circular shape at the center was considered based on previous investigations using a source point \citep{Bach2017} or cube tracer \citep{Seidel2022} released at the center of the cellbag \citep{Svay2020}. The degrees of mixing for different configurations, presented in Fig. \ref{Fig:tracer_ini}, exhibit significant differences. The center-circle and middle-line configurations show a faster increase in mixing compared to the other two cases. Furthermore, the left-half configuration shows a notable delay in mixing due to the symmetry of the geometry and rocking motion, as discussed in \citet{Gubanov2009}. The instantaneous tracer fields at the degrees of mixing of $\chi=0$ and $0.5$, illustrated in Fig. \ref{Fig:tracer_ini_snap}, highlight the dependency of the degree of mixing on initial tracer setups. While each tracer field has the same degree of mixing in each column, it exhibits different distributions, reflecting local flow structures. The top-half configuration primarily addresses mixing in the vertical direction, while the left-half configuration reflects mixing in the horizontal direction, indicating the entrainment of tracers on the left-hand side due to low mixing in that direction. The center-circle and middle-line configurations can also represent mixing in both vertical and horizontal directions. However, these latter two configurations are limited in their capacity to distribute over the entire domain due to their initial coverage being more isolated. Henceforth, we initiate tracers in the top half of the water phase to primarily probe vertical mixing. 

\subsection{Oxygen mass transfer coefficient} \label{subsec:oxygen}

\begin{figure}[t!]
\center \includegraphics[width=0.48\textwidth]{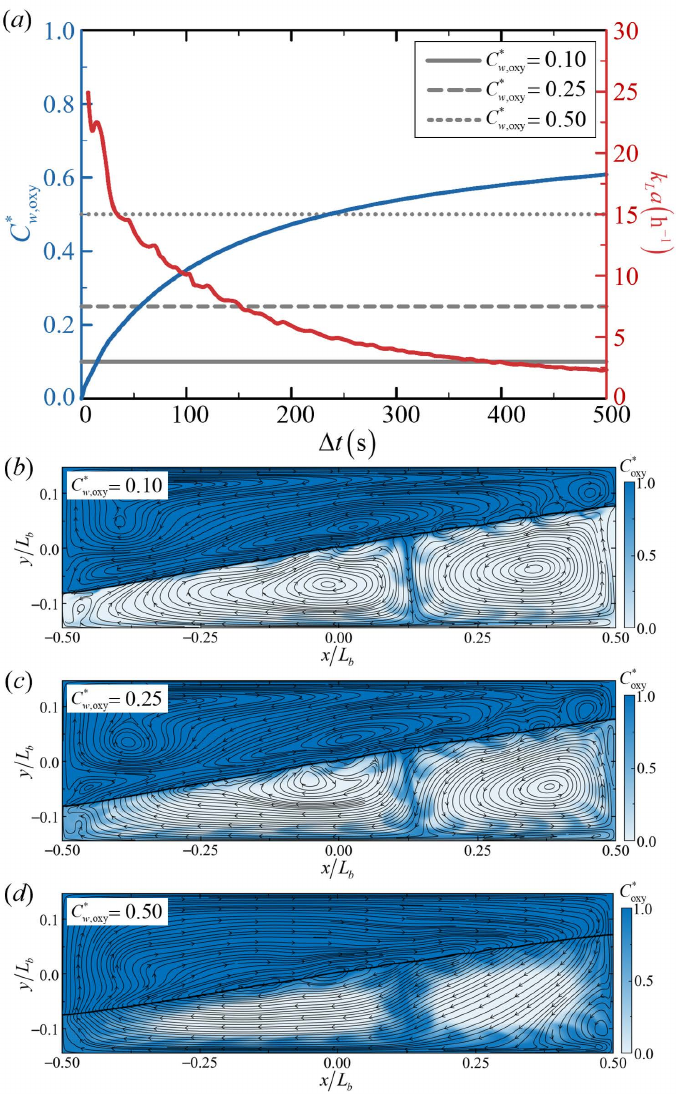}
\caption{(a) Time evolution of oxygen concentration in water (blue solid line), normalized by the oxygen saturation concentration, and (b) oxygen mass transfer coefficient (red solid line) for $\theta_{b,\textrm{max}}= 7^{\circ}$ and $f_b=32.5$ rpm. Snapshots of oxygen concentration at (b) $C_{w,\textrm{oxy}}^*=$ 0.10 (gray solid line), (c) 0.25 (gray dashed line), and (d) 0.50 (gray dotted line).} \label{Fig:oxy_time_evol}
\end{figure}

\begin{figure*}[t!]
\center \includegraphics[width=0.8\textwidth]{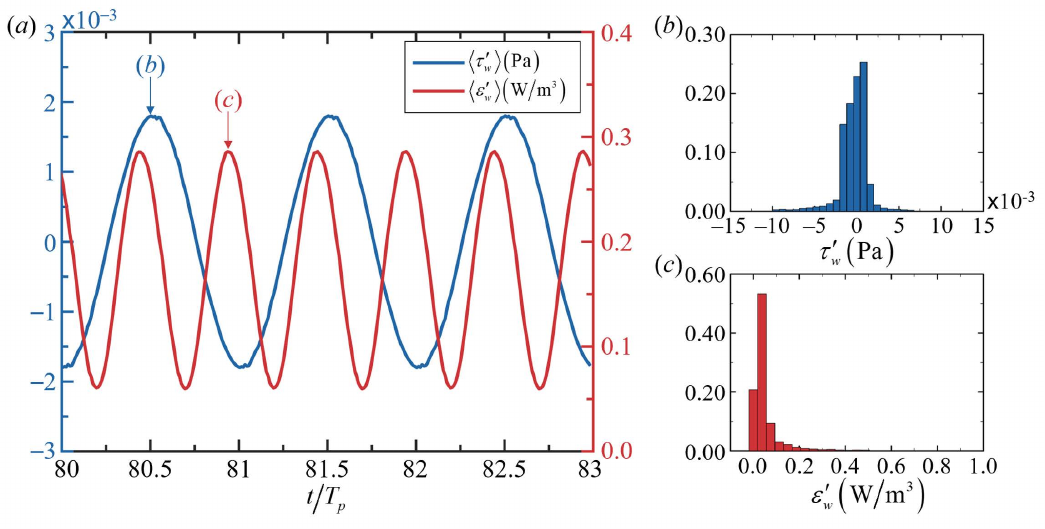}
\caption{(a) Time evolution of the spatially averaged shear stress (blue solid line) and energy dissipation rate (red solid line) in the non-inertial frame of reference for $\theta_{b,\textrm{max}}= 7^{\circ}$ and $f_b=32.5$ rpm. Normalized histograms of (b) shear stress and (c) energy dissipation rate distributions in water at times where the average is maximum values, as indicated in (a).} \label{Fig:tau_Ediss_evol}
\end{figure*}

The transport of oxygen is key for the growth of cultivated cells. In experiments, oxygen is supplied to the water through air delivered via a tube connected to the cell bag, and its concentration in the water is measured using a probe \citep{Junne2013,Bai2019a}. In our simulations, oxygen is replenished at each time step, providing the maximum possible oxygen transfer under the given operating parameters. Oxygen is introduced to the air side after 80 cycles, similar to the tracer case. Oxygen transfer across the water-air interface occurs through two mechanisms: diffusion and mass hold-up induced by wave breaking and interface roll-up \citep{Bai2019b}. In the laminar flow regime, oxygen mass hold-up is negligible, as it is primarily associated with turbulent flows. We thus only consider oxygen diffusion, governed by Henry's law near the interface. The total oxygen concentration in the water side is calculated as the sum over all grid cells: $\Sigma C_{w,\textrm{oxy}}$. The total saturation concentration in the water side is defined as: $C_\textrm{oxy,sat}=H_{\textrm{oxy}}\Sigma C_{a,\textrm{oxy}}$, where $C_{a,\textrm{oxy}}$ is the oxygen concentration in each air-side grid cell. The normalized oxygen level in the water phase, representing the degree of saturation, is given by $C_{w,\textrm{oxy}}^* = \Sigma C_{w,\textrm{oxy}}/C_{\textrm{oxy,sat}}$. The time evolution of this normalized oxygen level and snapshots at selected values are shown in Fig. \ref{Fig:oxy_time_evol}. Initially, the level of transferred oxygen increases rapidly due to a high concentration gradient near the interface (Fig. \ref{Fig:oxy_time_evol}(b)), but the transfer rate slows as oxygen accumulates near the interface (Fig. \ref{Fig:oxy_time_evol}(c) and (d)). Similar to the tracer case for vertical mixing (Fig. \ref{Fig:tracer_ini_snap}(b)), oxygen in the water tends to accumulate along regions near the boundaries of the vortical structures observed in the steady streaming, which are in contact with the air phase. The oxygen distribution is also aligned with the instantaneous vortical structure when the rocking amplitude reaches its maximum angle (Fig. \ref{Fig:vor_all}(b) and (d)), as illustrated in Fig. \ref{Fig:oxy_time_evol}(b).

The oxygen mass transfer coefficient, $k_La$, was computed based on the time evolution of $C_{w,\textrm{oxy}}^*$. In the classical $k_La$ approach, the flux of dissolved oxygen in the water is assumed to be proportional to the difference between the current oxygen level and the saturation level, akin to first-order kinetics. This simplifying relationship is represented by \citet{Singh1999}:
\begin{equation}
    \frac{dC_{w,\textrm{oxy}}^*(t)}{dt} = k_La(1-C_{w,\textrm{oxy}}^*(t)). \label{eqn:kla}
\end{equation}
By applying this relation locally to $C_{w,\textrm{oxy}}^*$, the instantaneous value of $k_La$, assumed to be locally constant around $t=t_0$, is given by:
\begin{equation} \label{eq:kla_fit}
    k_La(t_0) = -\frac{1}{\Delta t(t_0)} \mathrm{ln} \left(\frac{1-C_{w,\textrm{oxy}}^*(t)}{\beta_0(t_0)}\right),
\end{equation}
where $\beta_0(t_0)$ is a fitting coefficient evaluated at $t=t_0$. The value of $k_La$ is estimated using a moving window centered at $t_0$, by fitting the local values of $C_{w,\textrm{oxy}}^*$ to five consecutive time points within a time window of duration $\Delta t(t_0)$. Using five points yields results nearly identical to those obtained with a larger number of data points, as detailed in \ref{app:oxy_cal}, where different approaches for calculating $k_La$ are also compared. To examine temporal changes in $k_La$, we analyze values at $C_{w,\textrm{oxy}}^*=$ 0.1, 0.25, and 0.5.
 
Conventionally, $k_La$ has been estimated using Higbie's penetration theory \citep{Higbie1935}, with a properly defined hydrodynamic timescale in numerical simulations \citep{Farsoiya2021}. This approach has been predominantly applied in the turbulent flow regime \citep{Bai2019b,Seidel2022,Svay2020,Zhan2019}, where the Kolmogorov timescale is assumed to be the characteristic time scale. Based on this theory, these studies developed scaling formulations between $k_La$ and operating conditions, considering oxygen transfer across the interface and gas hold-up due to interfacial motion. Although the estimated $k_La$ provides a reasonable approximation for describing trends observed in experimental studies in some cases, our approach directly computes $k_La$ by considering the evolution of the full oxygen concentration field from first principles. This method is particularly valuable in the laminar flow regime, where no clear universal relationship exists between the hydrodynamics and oxygen transfer.

\subsection{Shear stress and energy dissipation rate} \label{subsec:shear}

Shear stress and energy dissipation rate (EDR) are commonly quantified to characterize potential cell damage in bioreactors \citep{Svay2020,Hu2011,Mcrae2024}. In particular, mammalian cells--often used for cultivated meat production--are highly sensitive to hydrodynamic shear in bioreactors. The shear stress in the water is defined as:
\begin{equation}
    \tau_w' = \mu_w \left(\frac{\partial u_x'}{\partial y'} + \frac{\partial u_y'}{\partial x'} \right) ,
\end{equation}
where $\mu_w$ is the viscosity of water. The EDR in the water is calculated as:
\begin{equation}
    \small
    \epsilon_w' = \mu_w \left[ 2\left(\frac{\partial u_x'}{\partial x'}\right)^2 + 2\left(\frac{\partial u_y'}{\partial y'}\right)^2 + \left(\frac{\partial u_x'}{\partial y'} + \frac{\partial u_y'}{\partial x'} \right)^2 \right].
\end{equation}
The spatially averaged shear stress, $\langle\tau_w'\rangle$ and EDR, $\langle\epsilon_w'\rangle$ in the water at $\theta_{b,\textrm{max}}= 7^{\circ}$ and $f_b=32.5$ rpm are presented in Fig. \ref{Fig:tau_Ediss_evol}(a). Both shear stress and EDR exhibit periodic behavior, while the EDR shows a phase shift relative to the rocking cycles. It is worth noting that the spatially averaged values of shear stress and EDR are lower than their local maximum values. Specifically, at the maximum values of shear stress and EDR, the normalized histogram of their distributions indicates that cells may experience locally high shear stress and EDR--up to five times greater than the mean values. (Fig. \ref{Fig:tau_Ediss_evol}(b) and (c)). However, the threshold EDRs inducing a lethal response in animal cells range from the order of $O(10^3)$ to $O(10^6)$ (W/m$^3$) \citep{Hu2011}, which are significantly higher than the EDR measured in the current simulations.


\section{Hydrodynamic impact on bioreactor performance} \label{sec:results}

In the following subsections, we investigate the mixing time, oxygen mass transfer coefficient, shear stress, and energy dissipation rate for different rocking frequencies and amplitudes to characterize how the hydrodynamic effects, including steady streaming and instantaneous flow structures, relate to these parameters. Videos of the volume fraction, vorticity, tracer field, and oxygen field for the baseline case ($\theta_{b,\textrm{max}}= 7^{\circ}$, $f_b=$ 32.5 rpm) are available in the supplementary material.

\subsection{Mixing time}\label{subsec:results-mixing}

\begin{figure}[t!]
\center \includegraphics[width=0.48\textwidth]{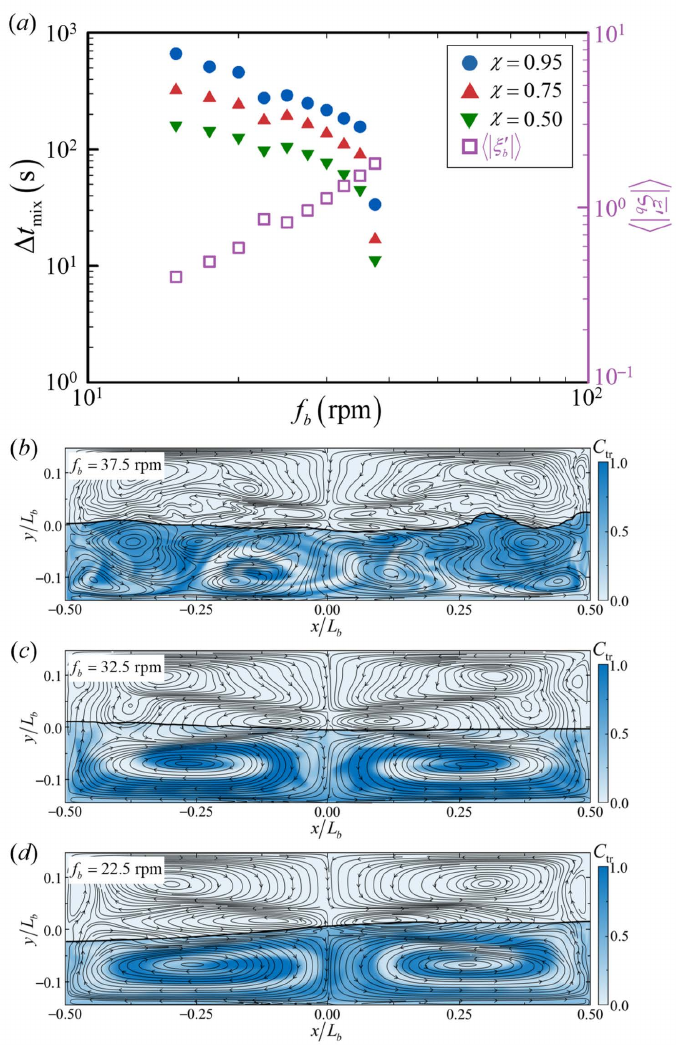}
\caption{Mixing at different rocking frequencies. (a) Mixing times at $\chi=$ 0.95 (solid blue circles), 0.75 (solid red upper triangles), and 0.50 (solid green lower triangles) and the spatially averaged absolute value of steady streaming vorticity in the water phase (hollow pink squares), for different rocking frequencies at $\theta_{b,\textrm{max}}= 7^{\circ}$. The left axis represents the mixing time, and the right axis represents the vorticity magnitude. Snapshots of tracer fields overlaid with streamlines of steady streaming for (b) $f_b=$ 37.5, (c) 32.5, and (d) 22.5 rpm, at the moment when $\chi = 0.5$ is achieved.} \label{Fig:dt_mix_rpm}
\end{figure}

Mixing in a rocking bioreactor strongly depends on the hydrodynamic characteristics of the flow for different rocking frequencies and amplitudes. To evaluate mixing, the degree of mixing is computed using Eq. \ref{eq:mixing_chi} with the top-half configuration of the initial tracer field, as described in Section \ref{subsec:mixing}. We first vary the rocking frequency while keeping the maximum rocking angle fixed at $\theta_{b,\textrm{max}} = 7^\circ$ (Fig. \ref{Fig:dt_mix_rpm}). Three different degrees of mixing, $\chi = $0.5, 0.75, and 0.95, are selected to analyze the evolution of mixing at different mixing stages (Fig. \ref{Fig:dt_mix_rpm}(a)). For all degrees of mixing, the mixing time $\Delta t_{\textrm{mix}}$ generally increases as the rocking frequency decreases. This trend is consistent with the strength of the steady streaming flow: the spatially averaged absolute value of steady streaming vorticity in the water phase $\langle |\xi_b'| \rangle$ increases with frequency, indicating a strong correlation between steady streaming and mixing time. Here, the steady streaming vorticity refers to the period-averaged instantaneous vorticity at each grid cell.
 
The instantaneous tracer distribution overlaid with the streamlines of steady streaming illustrates the relationship visually, as shown in Fig. \ref{Fig:dt_mix_rpm}(b--d). At $f_b=37.5$ rpm, both the tracer field and the streamlines of the averaged velocity fields exhibit an asymmetric and irregular distribution, indicating the emergence of turbulent behavior. This transition to turbulence significantly reduces mixing times (Fig. \ref{Fig:dt_mix_rpm}(a)), although it is associated with increased shear stresses, discussed in detail in Section \ref{subsec:shear}. In contrast, at the baseline case ($f_b=32.5$ rpm) and at lower frequencies, the instantaneous tracer fields align well with the steady streaming flow structure, consisting of two counter-rotating vortical structures that induce a roll-up behavior of the tracer fields (Fig. \ref{Fig:dt_mix_rpm}(c)). Surprisingly, at $f_b=22.5$ rpm, $\langle |\xi_b'| \rangle$ is higher, and mixing times are shorter than at neighboring frequencies, following a monotonic trend due to  interfacial motion caused by a nonlinear resonant behavior (Fig. \ref{Fig:dt_mix_rpm}(d)). This phenomenon is discussed and analyzed further in Section \ref{sec:spectral}.

\begin{figure}[t!]
\center \includegraphics[width=0.48\textwidth]{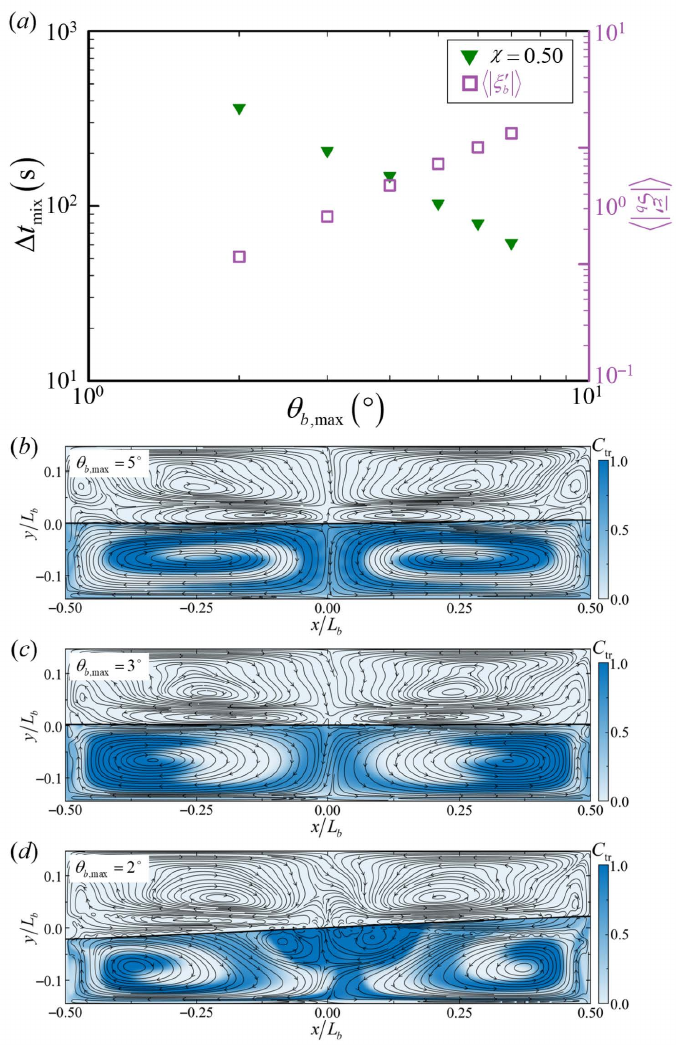}
\caption{Mixing at different rocking angles. (a) Mixing times at $\chi = 0.50$ (solid green lower triangles) and the spatially averaged absolute value of steady streaming vorticity in the water phase (hollow pink squares), for different rocking angles at $f_b=$ 32.5 rpm. The left axis represents the mixing time, and the right axis represents the vorticity magnitude. Snapshots of tracer fields overlaid with streamlines of steady streaming for (b) $\theta_{b,\textrm{max}}=$ $5^{\circ}$, (c) $3^{\circ}$, and (d) $2^{\circ}$ when $\chi = 0.5$ is achieved.} \label{Fig:dt_mix_deg}
\end{figure}

We next vary the maximum rocking angles while keeping the rocking frequency constant at $f_b=32.5$ rpm, as shown in Fig. \ref{Fig:dt_mix_deg}(a). The degree of mixing is evaluated here only at $\chi=0.5$ since mixing at lower rocking angles is significantly slower, requiring much longer simulation times. Similar to the case of varying rocking frequencies, the degree of mixing decreases monotonically with the maximum rocking angle due to a corresponding increase in $\langle |\xi_b'| \rangle$. This behavior is further illustrated through snapshots at different rocking angles in Fig. \ref{Fig:dt_mix_deg}(b--d). At all rocking angles, the instantaneous tracer fields closely follow the underlying steady streaming flow structure. However, as the rocking angle decreases, the size of vortical structures in the steady streaming diminishes, and their core locations shift toward the walls. This shift coincides with the emergence of additional counter-rotating structures near the centerline ($x/L_b=0$), which influence the tracer distribution and subsequent mixing time. 

It has been previously demonstrated that mixing time in a rocking bioreactor strongly depends on operating conditions \citep{Junne2013,Bartczak2022,Seidel2022}. 
Specifically, for a 2L Cytiva cellbag, the average mixing time--defined as the time to reach 95 $\%$ mixing--was reported at a maximum rocking angle of $7^\circ$ and rocking frequencies of $f_b =$ 2, 21, and 40 rpm \citep{Bartczak2022}. At $f_b =$ 40 rpm, the mixing time is approximately 4 seconds across different liquid volumes due to highly turbulent behavior, similar to our simulation results at $f_b=$ 37.5 rpm (Fig. \ref{Fig:dt_mix_rpm} (b)). However, at $f_b =$ 21 rpm, the mixing time increases to the order of $O(10)$ seconds. Similarly, at a lower rocking angle of $\theta_{b,\textrm{max}}=$ $2^{\circ}$, the mixing time remains on the order of $O(10)$ seconds at $f_b=$ 21 rpm \citep{Bartczak2022}. 
In the experiments by \citet{Eibl2009} with a 2L cellbag at a lower rocking angle of $\theta_{b,\textrm{max}}=$ $5^{\circ}$, the mixing time is approximately 80 seconds at $f_b=$ 16 rpm and 25 seconds at $f_b=$ 24 rpm, which deviates from our results at similar rocking angles (Fig. \ref{Fig:dt_mix_deg}(a)). This discrepancy is likely due to differences in measurement methodology and the initial distribution of the tracer. In their experiments, mixing time was estimated using pointwise pH probe measurements, which may introduce larger uncertainty at low rocking angles, where tracer material can remain locally trapped, as observed in Fig. \ref{Fig:dt_mix_deg}(b--d). Moreover, as discussed in Section \ref{subsec:mixing}, the mixing time is highly sensitive to the initial tracer distribution, which contributes to the differences between experimental measurements and our simulation results.

Previous studies have found that larger rocking angles and higher frequencies lead to shorter mixing times, primarily due to increased instantaneous velocity and vorticity, without note of steady streaming effects. Moreover, since these studies predominantly focused on the turbulent flow regime with a limited parameter space, further investigation is needed to understand laminar and transition regimes, as well as resonant behaviors--key aspects explored in the present study.

\subsection{Oxygen mass transfer coefficient} \label{subsec:results-oxy}

\begin{figure}[t!]
\center \includegraphics[width=0.48\textwidth]{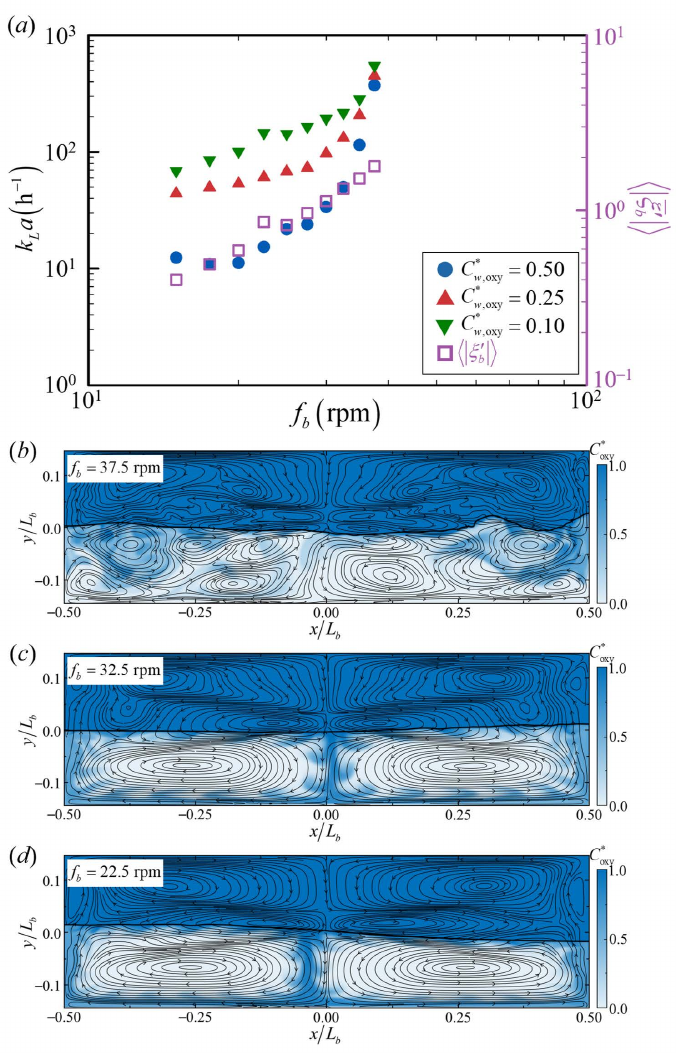}
\caption{Oxygen mass transfer coefficient at different rocking frequencies. (a) Oxygen mass transfer coefficient at normalized oxygen levels in the water of 0.50 (solid blue circles), 0.25 (solid red upper triangles), and 0.10 (solid green lower triangles) and the spatially averaged absolute value of steady streaming vorticity in the water phase (hollow pink squares), for different rocking frequencies at $\theta_{b,\mathrm{max}}= 7^{\circ}$. The left axis represents the oxygen mass transfer coefficient, and the right axis represents the vorticity magnitude. Snapshots of normalized oxygen concentration fields overlaid with streamlines of steady streaming for (b) $f_b=$ 37.5, (c) 32.5, and (d) 22.5 rpm when $C_{w,\textrm{oxy}}^*=$ 0.25 is achieved.} \label{Fig:kLa_rpm}
\end{figure}

Similar to the analysis of mixing time, we investigate the oxygen mass transfer coefficient under different operating conditions. The variation of $k_La$ with rocking frequency at a constant rocking angle $\theta_{b,\mathrm{max}}= 7^{\circ}$ is shown in Fig. \ref{Fig:kLa_rpm}. $k_La$ is computed using Eq. \ref{eq:kla_fit}, applying a moving window over five consecutive points from the time evolution of the normalized oxygen level. To examine the evolution of oxygen transfer at different stages, $k_La$ is calculated at $C_{w,\textrm{oxy}}^*=$ 0.10, 0.25, and 0.50.

Overall, $k_La$ increases with higher rocking frequencies, correlating with larger $\langle |\overline{\xi_b'}| \rangle$ (Fig. \ref{Fig:kLa_rpm}(a)). This suggests that oxygen transfer across the interface is enhanced at higher rocking frequencies, driven by the presence of stronger counter-rotating vortical structures in the steady streaming. Similar to the analysis of mixing time, at $f_b=37.5$ rpm, $k_La$ is significantly higher than in other cases due to the emergence of aperiodic turbulent-like behavior, as shown in Fig. \ref{Fig:kLa_rpm}(b). Since oxygen transfer occurs only through the interface, the transfer rate depends on the oxygen concentration gradient across it. Thus, local turbulent-like vortical structures replenish oxygen distribution near the interface more efficiently, significantly enhancing oxygen transfer.

In the baseline case ($f_b=$ 32.5 rpm) and at lower frequencies, the transferred oxygen tends to accumulate along the boundaries of steady streaming vortical structures. This behavior reduces the oxygen concentration gradient across the interface, thereby delaying oxygen transfer (Fig. \ref{Fig:kLa_rpm}(b--d)). At $C_{w,\textrm{oxy}}^*=$ 0.50, when half of the maximum saturated oxygen has already transferred into the water, $k_La$ plateaus at low frequencies ($f_b \leq$ 20 rpm) likely due to a similar oxygen concentration gradient near the interface being established over time. At $f_b=$ 22.5 rpm, $k_La$ at $C_{w,\textrm{oxy}}^*=$ 0.10 is higher than that at neighboring frequencies due to nonlinear interfacial motion--consistent with the elevated $\langle |\overline{\xi_b'}| \rangle$ at this frequency. However, this distinction diminishes at $C_{w,\textrm{oxy}}^*=$ 0.25 and 0.50.

\begin{figure}[t!]
\center \includegraphics[width=0.48\textwidth]{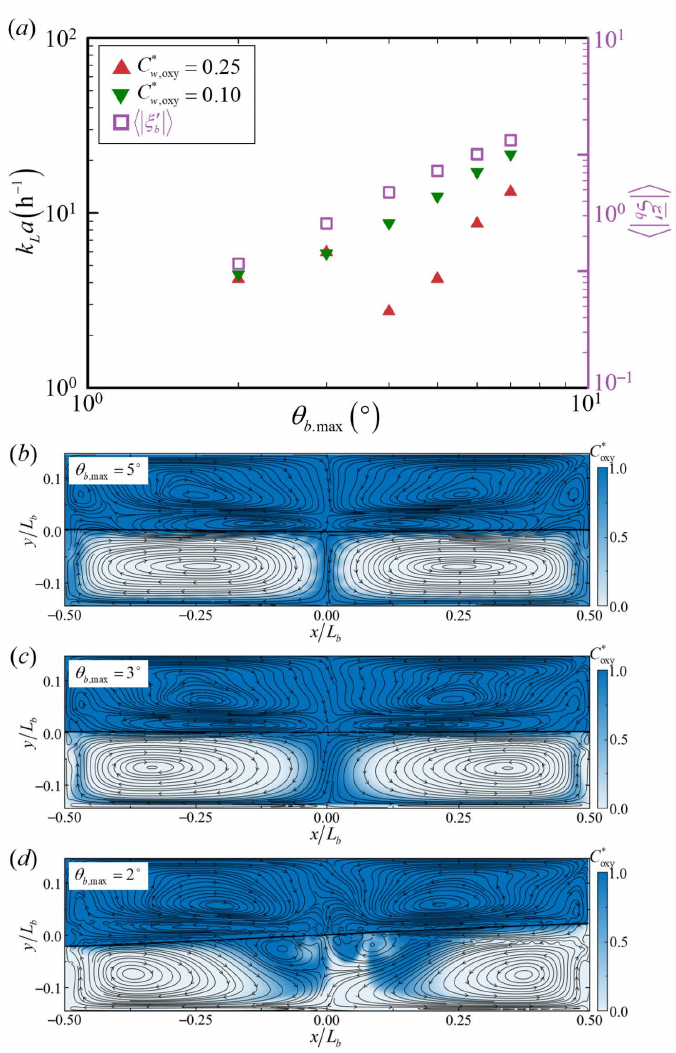}
\caption{Oxygen mass transfer coefficient at different rocking angles. (a) Oxygen mass transfer coefficient at normalized oxygen levels in the water of 0.25 (solid red upper triangles) and 0.10 (solid green lower triangles) and the spatially averaged absolute value of steady streaming vorticity in the water phase (hollow pink squares), for different rocking angles at $f_b=$ 32.5 rpm. The left axis represents the oxygen mass transfer coefficient, and the right axis represents the vorticity magnitude. Snapshots of normalized oxygen concentration fields overlaid with streamlines of steady streaming for (b) $\theta_{b,\textrm{max}}=$ $5^{\circ}$, (c) $3^{\circ}$, and (d) $2^{\circ}$ when $C_{w,\textrm{oxy}}^* = 0.25$ is achieved.} \label{Fig:kLa_deg}
\end{figure}

\begin{figure*}[t!]
\center \includegraphics[width=1.0\textwidth]{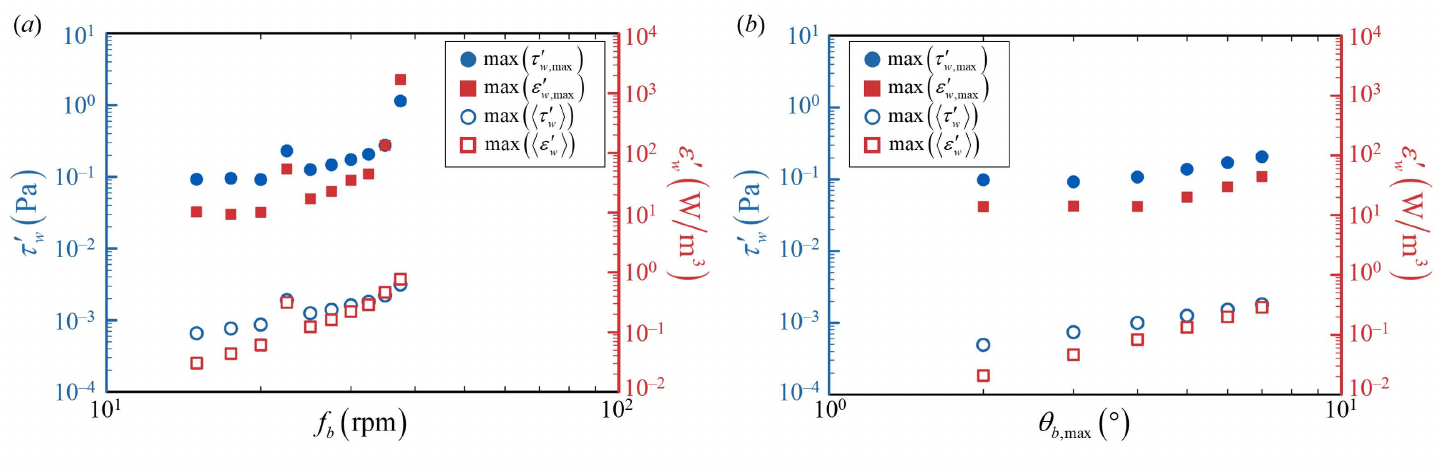}
\caption{Shear stress and energy dissipation rate at different rocking frequencies and angles. The maximum shear stress (solid blue circles) and energy dissipation rate (solid red squares) over time and in the water, along with the maximum of the spatially averaged shear stress (hollow blue circles) and energy dissipation rate (hollow red squares), are shown for (a) different rocking frequencies at $\theta_{b,\textrm{max}}= 7^{\circ}$ and (b) different rocking angles at $f_b=$ 32.5 rpm. The left axis represents the shear stress, and the right axis represents the energy dissipation rate.} \label{Fig:tau_Ediss_rpm_deg}
\end{figure*}

The oxygen mass transfer coefficient also depends on the maximum rocking angle at a fixed frequency of $f_b=32.5$ rpm (Fig. \ref{Fig:kLa_deg}(a)). In this case, $C_{w,\textrm{oxy}}^*=$ 0.10 and 0.25 are considered, as oxygen transfer is significantly delayed at smaller rocking angles, requiring longer simulation times. 

In the early stage ($C_{w,\textrm{oxy}}^*=$ 0.10), $k_La$ is lower at smaller rocking angles, corresponding to weaker steady streaming, as indicated by smaller values of $\langle |\overline{\xi_b'}| \rangle$. However, as more oxygen is transferred into the water ($C_{w,\textrm{oxy}}^*=$ 0.25), oxygen tends to accumulate along the boundaries of the counter-rotating vortical structures in the steady streaming. This accumulation reduces the oxygen concentration gradient near the interface and limits oxygen diffusion (Fig. \ref{Fig:kLa_deg}(b--d)). 

As the rocking angle decreases, the cores of the vortical structures shift closer to the sidewalls, and their strength and size diminish (Fig. \ref{Fig:kLa_deg}(c)). This shift again coincides with the formation of smaller counter-rotating vortical structures near the centerline ($x/L_b=0$) (Fig. \ref{Fig:kLa_deg}(d)). These additional small-scale vortical structures enhance oxygen transfer by replenishing local oxygen and increasing the concentration gradient at the interface near their locations. As a result, after the early stage--during which oxygen accumulation near the interface occurs--$k_La$ at lower rocking angles ($\theta_{b,\textrm{max}}=$ 2$^\circ$ and 3$^\circ$) becomes higher than at larger rocking angles when $C_{w,\textrm{oxy}}^*=$ 0.25.

Previous studies have investigated oxygen transfer in bioreactors under various operating conditions \citep{Eibl2003,Mikola2007,Junne2013,Zhan2019,Bai2019b,Svay2020}. In pointwise experimental measurements of dissolved oxygen in water, $k_La$ were reported to be on the order of $O(1)$ h$^{-1}$ for a half-filled 2L cellbag at $\theta_{b,\textrm{max}}=$ 6$^\circ$ and 10$^\circ$, with a rocking frequency of $f_b=$ 30 rpm \citep{Eibl2003}. These values increased for larger 20L and 200L cell bags. \citet{Junne2013} also reported $k_La$ of the same order of magnitude for a 10L cellbag, ranging from $O(1)$ h$^{-1}$ at $f_b=$ 15 rpm to $O(10)$ h$^{-1}$ at $f_b=$ 35 rpm. A simulation study using ANSYS CFX \citep{Svay2020} also reported comparable $k_La$ at a rocking frequency of $f_b=$ 40 rpm. These values are of the same order of magnitude as those obtained in our simulation results (Fig. \ref{Fig:kLa_rpm}(a)). The $k_La$ values obtained in the present study represent an upper bound of those measured experimentally, as the oxygen supply rate (or aeration rate) influences the $k_La$. In addition, in experiments, $k_La$ is typically evaluated when the measured oxygen concentration approaches saturation, whereas in our simulations, $k_La$ is computed at varying overall oxygen concentration levels. 

Most analyses in previous works rely on scaling relations for $k_La$ relevant in the turbulent flow regime. This relation is typically expressed in terms of flow characteristics (e.g., water density and viscosity, energy dissipation rate, and interface area) and operating conditions (e.g., rocking angle, frequency, and water volume). Most of these studies identified a monotonic behavior of $k_La$ across different rocking frequencies and angles, which is generally consistent with our simulation results. But, in the laminar flow regime, which is the focus of the present study, $k_La$ exhibits non-monotonic and highly dynamic behavior due to different steady streaming structures and resonance phenomena. These subtle effects can only be fully captured through simulations that couple oxygen transfer with hydrodynamic modeling, as in the current study.


\subsection{Shear stress and energy dissipation rate} \label{subsec:results-shear}

Rocking frequencies and maximum rocking angles influence shear stress and EDR in a rocking bioreactor (Fig. \ref{Fig:tau_Ediss_rpm_deg}). As the flow is unsteady, these quantities vary in both space and time.  Here, we compare two different measures of shear stress and EDR: one is the absolute maximum in water over one period (solid symbols in Fig. \ref{Fig:tau_Ediss_rpm_deg}), and the other is the maximum value of the spatially averaged quantity in water over one period (hollow symbols in Fig. \ref{Fig:tau_Ediss_rpm_deg}). It is worth noting that for flows showing turbulent-like behavior, these quantities are computed from 3,500 flow fields with a constant time gap of 0.05 simulation time. This corresponds to approximately 13 simulation points per cycle but is intentionally misaligned with the period to ensure convergence.

Overall, the maximum values of shear stress and EDR increase with higher rocking frequencies (Fig. \ref{Fig:tau_Ediss_rpm_deg}(a)). At $f_b=37.5$ rpm, where the flow transitions into the turbulent flow regime, the absolute maximum increases significantly, while the maximum of the spatially averaged values shows a more mild increase. At $f_b=22.5$ rpm, which corresponds to the resonance-like phenomena, all values are higher than those at neighboring frequencies. At lower frequencies ($f_b \leq 20$ rpm), the maxima of $\tau_{w,\textrm{max}}'$ and $\epsilon_{w,\textrm{max}}'$ reach a plateau, whereas the maxima of $\langle \tau_{w}' \rangle$ and $\langle \epsilon_{w}' \rangle$ are smaller. The local maximum of shear stress and EDR tend to occur near the interface \citep{Zhan2019}, showing similar behavior at lower frequencies, while their spatially averaged values decrease. The obtained mean shear stresses are of the same order of magnitude ($O(10^{-3})-O(10^{-2})$ Pa) as those reported in experiments and simulations \citep{Oncul2010} at $\theta_{b,\textrm{max}}=$ 7$^\circ$ and $f_b = 15$ rpm with a 2L cellbag. Similarly, higher maximum rocking angles generally lead to an increase in the maximum values of shear stress and EDR (Fig. \ref{Fig:tau_Ediss_rpm_deg}(b)). At low rocking angles ($\theta_{b,\textrm{max}}\leq$ 4$^\circ$), the maxima of $\tau_{w,\textrm{max}}'$ and $\epsilon_{w,\textrm{max}}'$ reach a plateau. 

These simulation results are consistent with the experimental findings reported by \citet{Oncul2010}, although only pointwise measurements were provided. Using a 2L cellbag, they found that at $\theta_{b,\mathrm{max}}=$ 7$^\circ$ and $f_b=$ 15 rpm, the maximum shear stress is on the order of $O(10^{-3})$ Pa, which is comparable to our results (Fig. \ref{Fig:tau_Ediss_rpm_deg}). The EDRs range from $O(10^2)$ to $O(10^3)$ W/m$^3$ for $\theta_{b,\mathrm{max}}=$ 7$^\circ$, falling between the local and spatially averaged maximum values observed in our simulations.

\begin{figure*}[t!]
\center \includegraphics[width=0.9\textwidth]{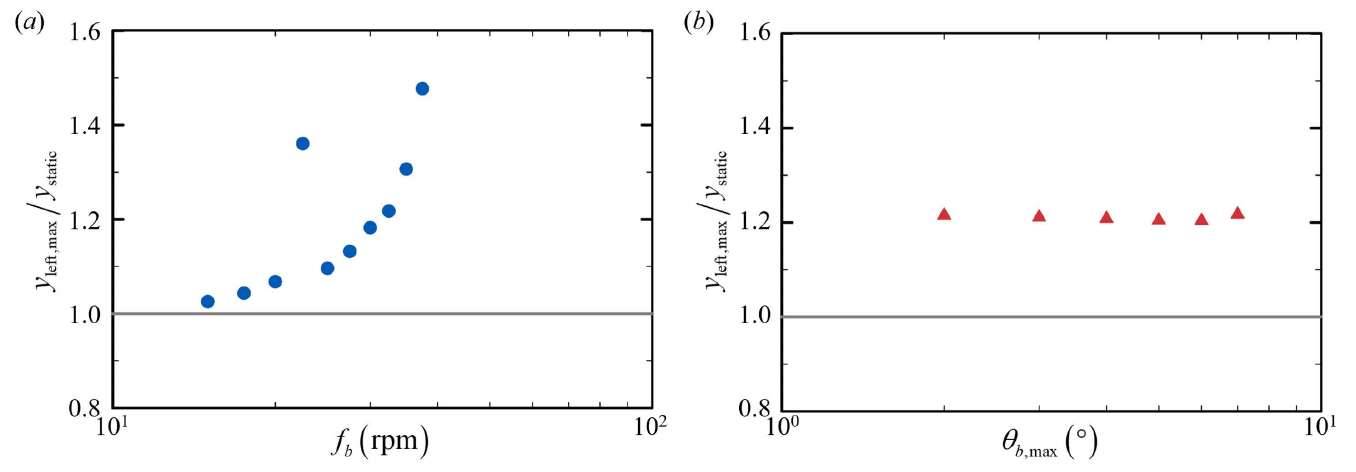}
\caption{Maximum surface elevation at different rocking frequencies and angles. The maximum surface elevations at the left end are normalized by those in the equivalent static scenario for (a) different rocking frequencies (solid blue circles) at $\theta_{b,\textrm{max}}= 7^{\circ}$ and for (b) different rocking angles (solid red triangles) at $f_b=$ 32.5 rpm. The static case is shown as gray lines.} \label{Fig:y_left}
\end{figure*}

The shear stress levels known to cause damage to mammalian cells range from 1 to 100 Pa \citep{Mardikar2000,Brindley2011,Haroon2022}. In our results, most maxima of $\tau_{w,\textrm{max}}'$ fall below this range, except in the transition regime at $f_b=37.5$ rpm and $\theta_{b,\textrm{max}}=$ 7$^\circ$. Recent studies have found that both excessively high and low shear stress can be more detrimental to cell viability than intermediate values (e.g., 10 Pa) \citep{Mardikar2000}. The effect of EDR on cell viability also varies depending on the cell type. For example, CHO-K1 cells, a widely used mammalian cell line for protein production, show a lethal response to EDR values on the order of $O(10^3)$ W/m$^3$ \citep{Hu2011}. In contrast, other cell types (e.g., hybridomia, MCF-7) have higher detrimental thresholds, ranging from $O(10^6)$ W/m$^3$ to $O(10^8)$ W/m$^3$ \citep{Hu2011,Mcrae2024}. In our results, most maxima of $\epsilon_{w,\textrm{max}}'$ remain well below these thresholds, except in the transition regime at $f_b=$ 37.5 rpm and $\theta_{b,\textrm{max}}=$ 7$^\circ$. Overall, most cases in the laminar flow regime show significantly lower shear stresses and EDRs, suggesting the potential feasibility of using a rocking bioreactor in the laminar regime for cell culture. However, for a more accurate assessment of cell viability, incorporating cell-hydrodynamics-coupled simulations would be needed for future studies.

\section{Interfacial motion} \label{sec:spectral}


Interface kinematics--such as interfacial motions and surface elevation--are closely linked to bioreactor characteristics \citep{Zhan2019}. Similar interfacial analyses have also been conducted in rocking systems with multiple phases (e.g., water tanks) to investigate the dynamical behavior of such systems and to predict potentially unstable conditions that could impact system safety 
\citep{Rocca2000,Ibrahim2005,Jung2015,Zhao2018}. In these studies, spectral analysis of interfacial motion in the frequency domain was used to identify distinct features of sloshing behavior under different geometries and forcing frequencies, particularly those relevant to resonant phenomena, which are likely associated with the hydrodynamic features of the rocking system. 

In the present study, we quantify the maximum surface elevation $y_\textrm{{left,max}}$ at different rocking frequencies and angles and compare them to the equivalent static case, where $y_\textrm{{left,max}} = 0.5L_b\theta_{b,\textrm{max}}$ (Fig. \ref{Fig:y_left}). Only the left-end elevation is considered because motions of interfaces in most cases are symmetric. For $\theta_{b,\textrm{max}} = 7^{\circ}$, the maximum surface elevation generally decreases as $f_b$ decreases (Fig. \ref{Fig:y_left}(a)), approaching one. However, similar to trends observed in bioreactor characteristics, the surface elevation becomes significantly larger at two distinct frequencies: $f_b = $37.5 and 22.5 rpm. The former corresponds to the transition to turbulent flow, while the latter is associated with resonance phenomena. For $f_b = $32.5 rpm, the surface elevation remains relatively consistent at $y_\textrm{{left,max}}/y_\textrm{{static}} \simeq 1.2$ for different $\theta_{b,\textrm{max}}$ (Fig. \ref{Fig:y_left}(b)).
 
\begin{figure*}[t!]
\center \includegraphics[width=0.9\textwidth]{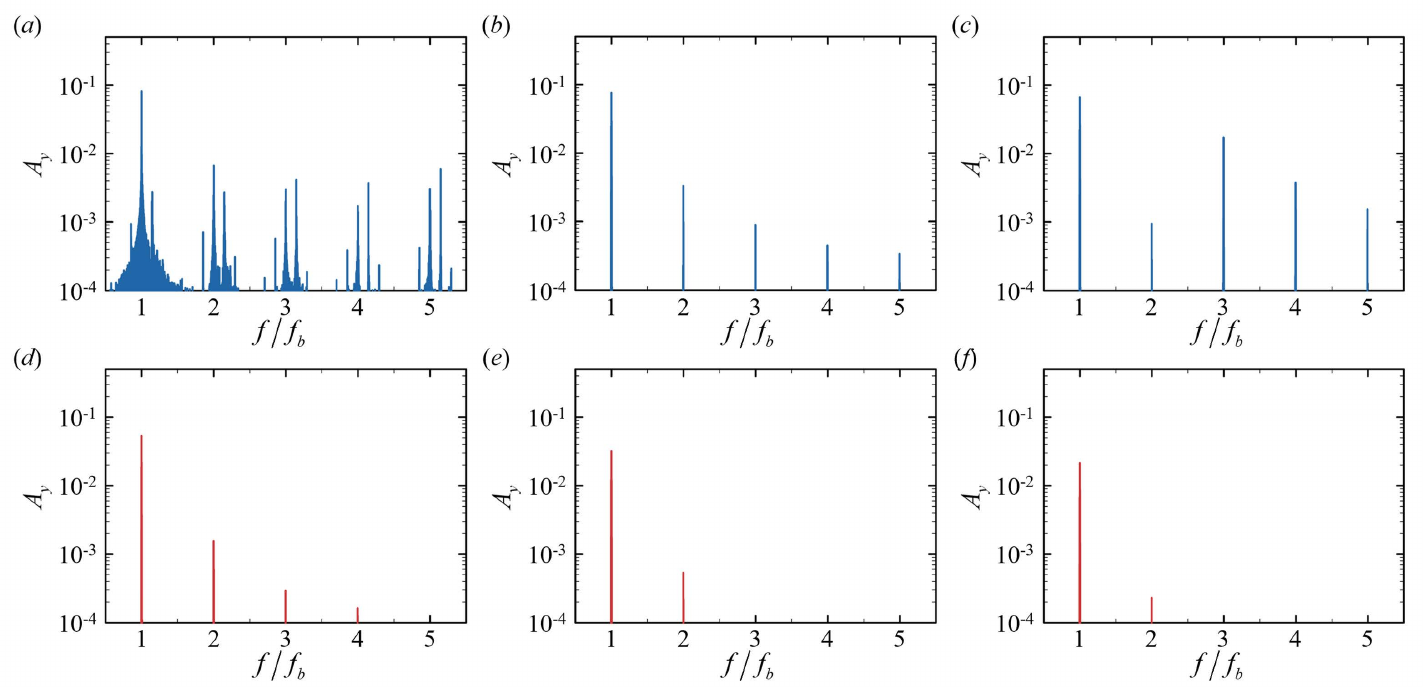}
\caption{Frequency spectra of surface elevation. Frequency spectra at different frequencies, normalized by the rocking frequency, for (a) $f_b=$ 37.5, (b) 32.5, and (c) 22.5 rpm at $\theta_{b,\textrm{max}}= 7^{\circ}$, and for (d) $\theta_{b,\textrm{max}}=$ $5^{\circ}$, (e) $3^{\circ}$, and (f) $2^{\circ}$ at $f_b=$ 32.5 rpm.} \label{Fig:FFT}
\end{figure*}

We then perform spectral analysis to characterize the harmonic behavior of the interfacial motion (Fig. \ref{Fig:FFT}). Following the analysis in Section \ref{sec:results}, we examine cases with $f_b=$ 37.5, 32.5, and 22.5 rpm at $\theta_{b,\textrm{max}} = 7^{\circ}$ (Fig. \ref{Fig:FFT}(a--c)) and cases with $\theta_{b,\textrm{max}} = 5^{\circ}$, $3^{\circ}$, and $2^{\circ}$ at $f_b=$ 32.5 rpm (Fig. \ref{Fig:FFT}(d--f)). As expected from the maximum surface elevation analysis (Fig. \ref{Fig:y_left}), the frequency spectra at $f_b=$ 37.5 rpm and $\theta_{b,\textrm{max}} = 7^{\circ}$ exhibit strong fluctuations across a broad frequency range, indicating the presence of turbulent-like behavior at the interface and confirming that this case falls within the transition regime. At lower rocking frequencies, the spectra peak only at integer multiples of the rocking frequency ($f=nf_b$, $n=1,2,3 ...$). For the baseline case ($f_b = 32.5$ rpm), the largest peak occurs at $f=f_b$, with the higher harmonics monotonically decreasing. However, at $f_b$ = 22.5 rpm, we see an anomalous amplification of the third harmonic ($f=3f_b$). These distinct spectral characteristics also correspond to distinct behaviors in the mixing time (Fig. \ref{Fig:dt_mix_rpm}(a)), $k_La$ (Fig. \ref{Fig:kLa_rpm}(a)), and shear stress and EDR (Fig. \ref{Fig:tau_Ediss_rpm_deg}) at this rocking frequency, distinguishing it from neighboring rocking frequencies. For $f_b = 32.5$ rpm, the contribution of the higher harmonics decreases as the rocking angle is decreased, suggestive of a more linear response. 
 
Resonant behavior of a rocking geometry has been observed in both a rocking bioreactor \citep{Zhan2019} and a sloshing rectangular water tank \citep{Rocca2000}. Linear resonant behavior occurs when an oscillatory system is driven at one of its natural frequencies. However, subharmonic and superharmonic resonances are possible in nonlinear systems, such as for the fluid system considered here. \citet{Zhan2019} conducted simulations using ANSYS-FLUENT with realistic cellbag geometries and found that flow velocity, mixing, and shear stress were enhanced at a rocking frequency of 15 rpm compared to higher frequencies (22 and 30 rpm). This behavior is likely related to the estimated first-mode natural frequency of the cellbag. Similarly, in the study of a rectangular water tank, \citet{Rocca2000} observed spectra peaks in surface elevation through both numerical simulations and experiments. These peaks appear at multiples of the driving frequency. Among them, the third harmonic exhibited the greatest amplitude, aligning closely with the natural frequency $f_n$ of the second wave mode ($n=2$). This behavior implies that the nonlinear resonance mechanism can cause the growth of the mode, enhancing interfacial motions. In our spectral analysis, a similar amplification of subharmonics at a rocking frequency of 22.5 rpm is observed, indicating the onset of nonlinearity driven by the unstable nature of the fluid near the natural frequency of the bioreactor.

The natural frequency of an incompressible, inviscid, and irrotational fluid with surface tension, in response to small perturbations in velocity potential and surface elevation within a rectangular geometry, is given by \citet{Ibrahim2005}:
\begin{equation}
 f_n = \frac{1}{2\pi}\sqrt{\left(gk_n + \frac{\sigma}{\rho_w}k_n^3\right) \rm{tanh}\it (k_nh_b)},   
\end{equation}
where $k_n = \pi n/L_b$ and the water depth is $h_b = 0.5L_b$. For our parameters, the first natural frequency is estimated to occur at $ f_1=$68.8 rpm, approximately equal to thrice the forcing frequency when $f_b=22.5$ rpm. As discussed throughout, this frequency is where we noted anomalous signatures in the mixing time, $k_La$, shear stress, and EDR, as well as interface kinematics.


\section{Discussion and concluding remarks}

We develop a computational framework using the Basilisk open-source platform to characterize a rocking bioreactor for cultivated meat production, focusing on mixing, oxygen transfer, shear stress, and energy dissipation rate. This framework employs a non-inertial frame of reference to model the rocking motion, allowing for fixed, non-moving computational boundaries, which simplifies the simulation setup and reduces computational cost. In addition, we solve the advection-diffusion equations while accounting for sharp concentration gradients across the interface, enabling tracer motion simulations in water with quantifiably negligible leakage into the air phase. Oxygen transfer from the air into the liquid is modeled based on Henry's law at the interface. To assess the potential impact on cell survival, we quantify local shear stress and energy dissipation rate. We also evaluate different approaches for calculating the degree of mixing, which we find to strongly depend on the initial tracer distribution, and the oxygen transfer mass coefficient, as determined by changes in oxygen concentration in water over a local time interval.

Using this framework, we investigate the relationship between hydrodynamics and mixing time, oxygen transfer mass coefficient, shear stress, and energy dissipation rate under different operating conditions. In general, as rocking frequency and maximum rocking angle increase, mixing and oxygen transfer are enhanced, accompanied by a rise in local shear stress and energy dissipation rate. In the laminar flow regime, where the flow exhibits periodic behavior, steady streaming--which induces net displacement of tracers and oxygen over a cycle--shows a consistent relationship with mixing time and the oxygen mass transfer coefficient. However, all quantities of interest exhibit distinct behaviors at two specific rocking frequencies for $\theta_{b,\textrm{max}}=7^\circ$. At rocking frequencies of $f_b=$ 37.5 rpm and 22.5 rpm, mixing time decreases, oxygen mass transfer coefficient increases, and shear stress and energy dissipation rate increase. However, the magnitude of these changes is significantly larger at 37.5 rpm than at 22.5 rpm. At $f_b=$ 37.5 rpm, the flow transitions into a turbulent-like aperiodic regime, corroborated by the distributed frequency spectra of surface elevation. Meanwhile, at $f_b=$ 22.5 rpm, a resonance mechanism--also reflected in the frequency spectra--induces distinct bioreactor performance. 

Our results reveal a close relationship between hydrodynamic features and the performance of a bioreactor. In the laminar regime, rocking frequency and maximum rocking angle are key parameters that govern the strength of the steady streaming, which primarily drives advective transport on long time scales. In previous studies, it has been demonstrated that the velocity of steady streaming exhibits a linear relationship with the oscillation amplitude and frequency of the container \citep{Bouvard2017}. This relationship was further explored theoretically \citep{Bongarzone2024} and validated in open-flow mixers \citep{Rallabandi2017} and stirred-type bioreactors \citep{Kahouadji2022}. Building upon these efforts, our computational framework enables more accurate quantification of mixing time and oxygen mass transfer coefficient by directly simulating tracer motion and oxygen transfer across an interface from first principles. We then relate the characteristics of steady streaming to these performance metrics.

Although rocking frequency and maximum rocking angle are key parameters, other factors also influence bioreactor performance. The filled water volume alters hydrodynamic behavior, enhancing or suppressing mixing and oxygen transfer. For smaller water volumes, mixing efficiency increases as the interaction between the bottom wall and the interface becomes more pronounced, which effectively controls the swirling direction of small-scale vortices \citep{Watanabe2022}. In addition, the aspect ratio of the cellbag plays a crucial role in overall bioreactor performance, particularly in oxygen transfer rates. Even for cellbags of the same volume, those with higher aspect ratios expose oxygen to a larger interfacial area, enhancing oxygen transfer \citep{Svay2020,Seidel2022}. Furthermore, the geometric details of actual cellbags influence the hydrodynamics of the bioreactor \citep{Oncul2010,Junne2013,Bai2019a}. These geometries are expected to affect the structure and shape of steady streaming, and in turn, the overall bioreactor performance. The number of degrees of freedom (DOF) of the bioreactor motion is another key operating factor. Although similar trends are observed in velocity profiles at different rocking frequencies when compared to single-DOF cases \citep{Zhan2019}, two-DOF motion generally improves mixing efficiency and reduces mixing time \citep{Seidel2022}. A more comprehensive study exploring these geometric and operating conditions is needed in future research.

The two-dimensional geometry studied herein effectively captures key hydrodynamic features of a rocking bioreactor, as its motion is predominantly in the vertical plane, even when considering a real-shaped cellbag \citep{Zhan2019}. While three-dimensional cellbags may induce weak fluid motion perpendicular to the vertical plane, the key dynamics remain well-represented in 2D simulations. Moreover, 2D and 3D simulations exhibit similar peaks in their frequency spectra of surface elevation \citep{Rocca2000}, suggesting comparable hydrodynamic characteristics, including resonant behavior. In our comparison of two-dimensional and three-dimensional simulations using the current cellbag geometry, flow characteristics show good agreement, as detailed in \ref{app:2D_3D}.

Beyond geometric and operating conditions, evaluation approaches are important for robust characterization of the bioreactor. Our current methods, which utilize tracer and oxygen concentration, provide a more accurate physics-based assessment of mixing and oxygen transfer compared to previous experimental \citep{Singh1999,Bai2019a,Bartczak2022} and simulation \citep{Svay2020,Seidel2022} studies. However, as discussed in Section \ref{subsec:mixing}, the initial tracer distribution significantly affects the time evolution of mixing degree and the corresponding mixing time. Recent efforts have sought to develop more robust evaluation methods that minimize dependence on initial tracer distribution and provide a more reliable measure of mixing efficiency \citep{Gubanov2009,Gleeson2005,Farazmand2017,Shi2024}. These approaches could play an important role in bioreactor optimization, which remains a key future research direction. Furthermore, for accurate evaluation of oxygen supply to cells, the mixing level of oxygen in water must be quantified. This metric would capture the oxygen distribution in water and help predict oxygen availability for cells. Our current setup assumes oxygen is initially uniformly distributed throughout the entire air phase, which reflects real-world conditions over short timescales. However, a more realistic long-term simulation would incorporate oxygen supply through a port in the cellbag, mimicking real cell culturing conditions. 

Our study demonstrates that local shear stresses and energy dissipation rates remain well below the lethal threshold for animal cells in both the laminar and transition regimes within our model system. However, a more comprehensive characterization of bioreactor performance and its impact on cell culturing is necessary. A key next step is to incorporate cell dynamics, including growth and death, using agent-based modeling \citep{Cantarero2024} or continuum biomass modeling \citep{Chao2015}. 

The developed computational framework provides a robust foundation for expanding functionality to broader aspects of bioreactor and cell culturing analysis, offering significant opportunities for future application in industrial settings. For example, practical cell culture systems often use non-Newtonian media \citep{Wyma2018,Zhang2025}, which should be incorporated into future extensions of the framework. Additionally, the numerical treatment of interface-boundary interactions should be improved to enable more accurate simulations across a wide range of rocking angles and frequencies. In particular, at large rocking angles, the interface may contact the top or bottom boundary, necessitating the implementation of dynamic contact angle models. Furthermore, complex interfacial phenomena such as wave breaking, gas holdup, and their effects on oxygen transfer at large rocking angles require further investigation for accurate representation. In the current setting, the framework solves equations for hydrodynamics, tracers, and oxygen transport simultaneously, which can be computationally intensive. For practical applications in the laminar flow regime, the hydrodynamic flow fields over one cycle could be precomputed and reused, allowing tracer and oxygen transport to be simulated on top of these flow fields. This approach would significantly reduce computational cost.

This work includes the careful development and public release of a code repository to support community use and further development. The released framework is intended to serve as a user-friendly tool for extending toward additional multi-physics elements, geometrical features, and operational regimes. This advancement may enable more efficient design and scale-up of bioreactors, ultimately contributing to the industrial production of cultivated meat.

\section*{CRediT authorship contribution statement}

\textbf{Minki Kim:} Investigation, Methodology, Validation, Visualization, Writing--original draft.
\textbf{Daniel M. Harris}: Conceptualization, Investigation, Methodology, Writing--review and editing, Supervision, Project administration.
\textbf{Radu Cimpeanu}: Conceptualization, Investigation, Methodology, Writing--review and editing, Supervision, Project administration.

\section*{Declaration of competing interest}
The authors declare that they have no known competing financial interests or personal relationships that could have appeared to influence the work reported in this paper.

\section*{Data availability}
The codes used for bioreactor simulations are publicly available at: \href{https://github.com/rcsc-group/BioReactor}{https://github.com/rcsc-group/BioReactor}.

\section*{Acknowledgments}
The authors gratefully acknowledge the financial support of the Good Food Institute, the Cultivated Meat Modeling Consortium, and the United States Department of Agriculture (USDA) National Institute of Food and Agriculture at Tufts University. This research used computational resources and services at the Center for Computation and Visualization, Brown University, and Bridges-2 at Pittsburgh Supercomputing Center through allocation AGR240019 from the Advanced Cyberinfrastructure Coordination Ecosystem: Services and Support (ACCESS) program, which is supported by National Science Foundation grants \#2138259, \#2138286, \#2138307, \#2137603, and \#2138296. The authors would also like to acknowledge Benny Smith, Emma Abele, and Ajay Harishankar Kumar for their preliminary work, and Glenn R. Gaudette, Luke R. Perreault, Elvis Alexander Agüero Vera, and Luke Rossi for valuable discussions and feedback.

\appendix

\section{Determination of a simulation resolution} \label{app:grid}

\begin{figure*}[t!]
\center \includegraphics[width=1.0\textwidth]{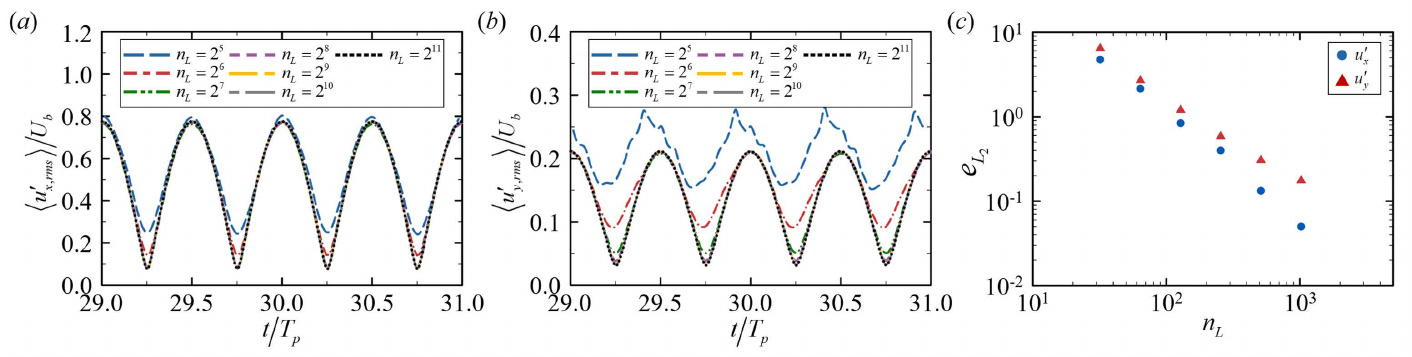}
\caption{Convergence of velocity components and their errors at different grid resolution levels. Time evolution of root-mean-square values of (a) $x'$- and (b) $y'$-velocity components on the liquid side in the non-inertial frame of reference over two cycles for different grid resolutions, ranging from $n_L=2^5$ to $2^{11}$ computational cells per dimension. (c) $L_2$ norms of the velocity components for different resolution levels.} \label{Fig:app_resol}
\end{figure*}

The hydrodynamics of the bioreactor at different grid resolution levels are investigated to verify convergence. In the baseline case ($\theta_b=7^\circ$ and $f_b=32.5$ rpm), the root-mean-square (rms) values of the liquid-phase velocity components in the non-inertial frame of reference, $u_{x,rms}'$ and $u_{y,rms}'$, are examined for grid resolutions ranging from $n_L=2^5$ to $n_L=2^{11}$ cells, where $n_L$ is the number of cells along the reference length scale (Fig. \ref{Fig:app_resol}(a) and (b)). For low resolutions ($n_L=2^5-2^7$), the velocity components vary significantly with resolution. However, as $n_L$ increases, they converge toward the highest resolution case ($n_L=2^{11}$). The numerical errors in these velocity components are evaluated at $t/T_p=29.77$, when the rocking angle reaches the maximum. The discrete $L_2$ norm of the error for each velocity component is computed as: 
\begin{equation}
    e_{i,L_2} = \left(\frac{1}{N_e}\sum_{e=1}^{N_e}\left(u_i' - u_{i,\mathrm{ref}}' \right)^2 \right)^{1/2}\bigg/\langle u_{i,\mathrm{ref}}'\rangle ,
\end{equation}
where $i=x$ and $y$, $N_e$ is the number of cells, and the subscript ref represents the reference case with $n_L=2^{11}$. The errors at different resolution levels are presented in Fig. \ref{Fig:app_resol}(c). The errors for both velocity components exhibit good convergence, decreasing consistently as $n_L$ increases.

\section{Calculation of oxygen mass transfer coefficient} \label{app:oxy_cal}

\begin{figure}[t!]
\center \includegraphics[width=0.48\textwidth]{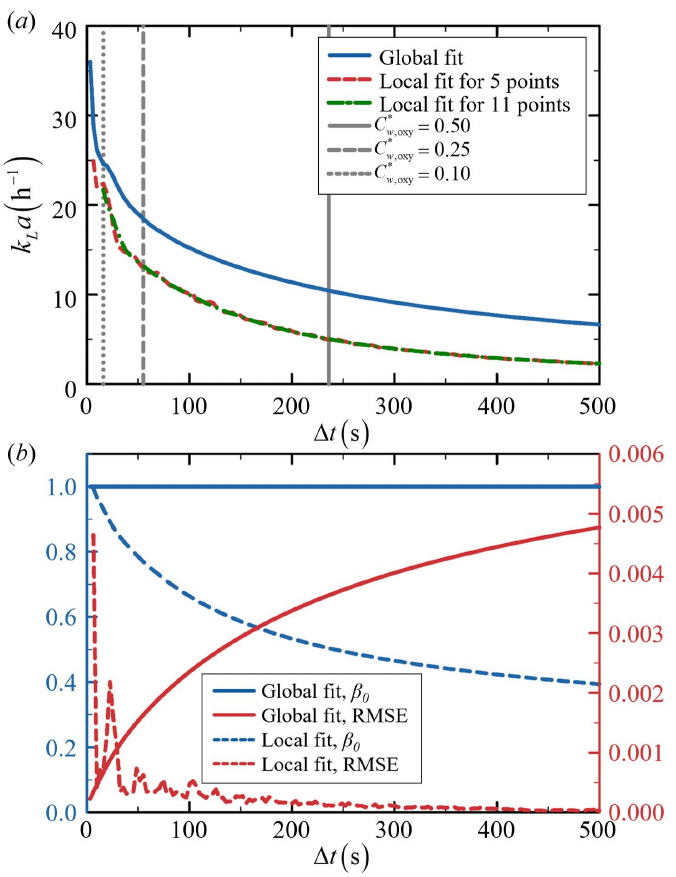}
\caption{(a) Time evolution of the oxygen mass transfer coefficient calculated using different methods: global fit (blue solid line) and local fits with 5 (red dashed line) and 11 (green dashed-dot line) consecutive points. Different oxygen concentrations, $C_{w,\textrm{oxy}}^*=$ 0.5,0.25, and 0.1, are indicated in (a). (b) $\beta_0$ (left axis) and RMSE (right axis) of the global fit and the local fit with 5 consecutive points.} \label{Fig:app_oxy_fit}
\end{figure}

We compare the oxygen mass transfer coefficients computed using different approaches, all based on the time evolution of $C_{w,\textrm{oxy}}^*$ (Fig. \ref{Fig:app_oxy_fit}). Appealing to the solution of Eq. \ref{eqn:kla}, where $k_La$ is assumed constant, the analytical expression for $k_La$ over a measurement time interval $[t_1, t_2]$ is given by:
\begin{equation}
    k_La = \left(\frac{1}{t_2-t_1}\right) \mathrm{ln}\left[\frac{1-C_{w,\textrm{oxy}}^*(t_1)}{1-C_{w,\textrm{oxy}}^*(t_2)}\right].
\end{equation}
By fitting the oxygen concentration globally from the initial time ($t = 0$) to the current time $t_0$, $k_La$ can be calculated as:
\begin{equation}
    k_La(t_0) = -\left(\frac{1}{t_0}\right) \mathrm{ln}\left[1-C_{w,\textrm{oxy}}^*(t_0)\right],
\end{equation}
which corresponds to the case of Eq. \ref{eq:kla_fit} with $\beta_0=$ 1. By reducing the fitting window to a small number of local data points, a localized $k_La$ estimate can be obtained using Eq. \ref{eq:kla_fit}.

$k_La$ computed using the global fit and local fits with 5 and 11 consecutive points is compared in Fig. \ref{Fig:app_oxy_fit}(a). $\beta_0$ and the root-mean-squared error (RMSE) for each approach are introduced in Fig. \ref{Fig:app_oxy_fit}(b). Both approaches show that $k_La$ decreases over time due to the slowing oxygen transfer rate. However, $k_La$ computed using the local fit is consistently lower than that from the global fit. The RMSE for the global fit is an order of magnitude higher than that of the local fit, indicating better accuracy for the local fit. Thus, in Sections \ref{subsec:oxygen} and \ref{subsec:results-oxy}, we quantify $k_La$ for different rocking frequencies and angles using the local fit with five consecutive data points.

\section{Effect of bioreactor dimensions on flow characteristics} \label{app:2D_3D}

\begin{figure}[t!]
\center \includegraphics[width=0.48\textwidth]{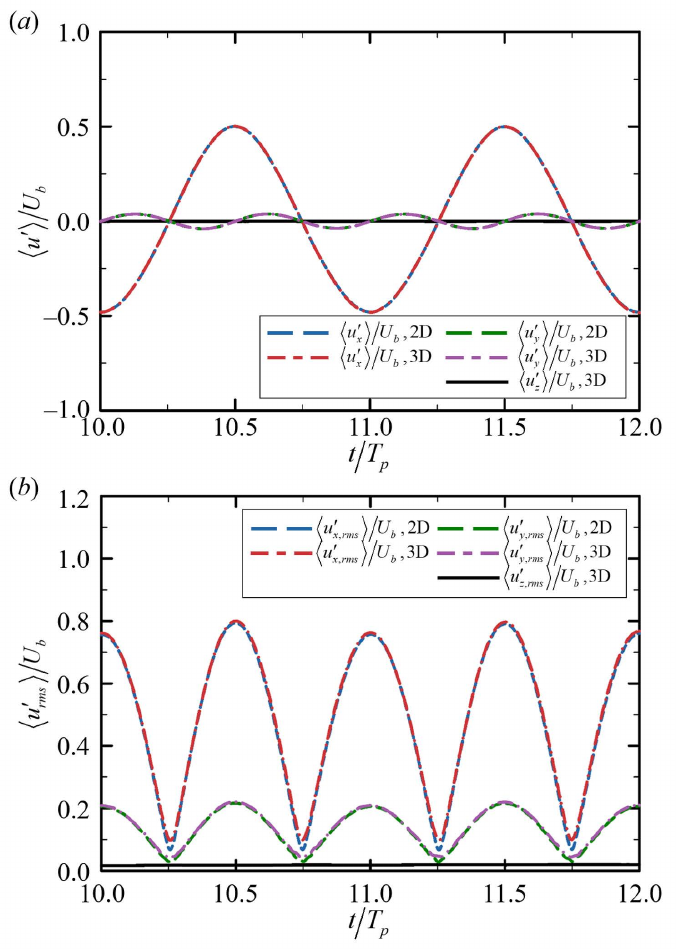}
\caption{Comparison between two-dimensional (2D) and three-dimensional (3D) simulations. Time evolution of (a) spatially averaged values and (b) rms values of the $x'$-, $y'$-, and $z'$-velocity components on the liquid side in the non-inertial frame of reference over two cycles for 2D and 3D simulations. The rocking conditions are $\theta_{b,\textrm{max}}= 7^{\circ}$ and $f_b=$ 32.5 rpm. Each velocity component is normalized by the characteristic velocity scale.} \label{Fig:app_2D_3D}
\end{figure}

Three-dimensional (3D) simulations are conducted to validate the two-dimensional (2D) simplification of a bioreactor geometry. The spatially averaged and rms values of each velocity component on the liquid side at $\theta_{b,\textrm{max}}= 7^{\circ}$ and $f_b=$ 32.5 rpm are presented in Fig. \ref{Fig:app_2D_3D}. For simplicity, the lengths in the $y$ and $z$ directions are set equal in the 3D simulation, which also reflects the typical dimensional ratios used in cellbags. The grid resolution is $n_L=2^9$ cells per characteristic length scale for both 2D and 3D cases. To reduce computational load, regular oscillation is achieved after 10 seconds. Velocities are compared after 10 cycles. 

Both the spatially-averaged values (Fig. \ref{Fig:app_2D_3D}(a)) and rms values (Fig. \ref{Fig:app_2D_3D}(b)) of the $x$ and $y$-velocity components in the 2D and 3D cases show good agreement. The $z$-velocity exhibits significantly smaller spatially averaged and rms values compared to the other components, confirming the suitability of 2D simulations for representing the dominant flow behavior in the 3D case. Similar consistency has also been observed in \citet{Zhan2019} using ANSYS FLUENT at $\theta_{b,\textrm{max}}=$ 7$^\circ$ and $f_b=$ 30 rpm. 

\bibliographystyle{elsarticle-harv} 
\bibliography{IJMF_bio}


\end{document}